\documentclass[11pt, twoside]{report}
\usepackage[a4paper, left=3cm, right=3cm, top=3.5cm, bottom=4.5cm]{geometry}
\usepackage{xcolor}
\usepackage{url}
\usepackage{graphicx}
\usepackage{longtable}
\usepackage[font=small]{caption}
\usepackage{subcaption}
\usepackage{booktabs}
\usepackage{multirow}
\usepackage{csquotes}
\usepackage{array}
\usepackage{mathtools}
\usepackage{amsmath}
\usepackage{amsthm}
\usepackage{amssymb}
\usepackage{algorithm}
\usepackage[noend]{algpseudocode}
\usepackage{fancyhdr}

\DeclareMathOperator*{\argmax}{arg\,max}
\DeclareMathOperator*{\argmin}{arg\,min}

\theoremstyle{definition}
\newtheorem{definition}{Definition}[section]

\author{Thomas Vecchiato}
\date{October 2024}

\begin{document}

\begin{titlepage}
    \newgeometry{top=1.25cm,bottom=2.5cm,right=2cm,left=2cm}
    \includegraphics[width=0.2\textwidth]{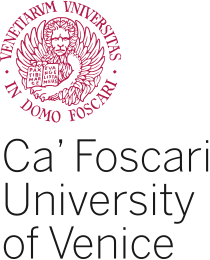}
    
    \begin{center}
        \vspace{1.5cm}
        \large
        Master's Degree Programme in\\
        Computer Science and Information Technology, CM90 \\
        \vspace{0.5cm}
        Curriculum \\
        Artificial Intelligence and Data Engineering

        \vspace{2.5cm}
        \textit{Final Master Thesis}

        \vspace*{0.75cm}
        \huge
        \textbf{Learning Cluster Representatives for Approximate Nearest Neighbor Search}
        
        \vspace{0.75cm}
        \large
        \textit{by}

        \vspace{0.5cm}
        \LARGE
        \textbf{Thomas Vecchiato} \\
        \large
        Student Number 880038

        \vspace{2.75cm}
        Supervisor
        
        Prof. Claudio Lucchese

        \small
        Ca' Foscari University of Venice
        
        \vspace{0.5cm}
        \large
        Co-Supervisor
        
        Dr. Sebastian Bruch
        
        \small
        Northeastern University
    
        \vspace{1.25cm}
        \large
        Academic Year
        
        2023/2024
        
    \end{center}

\end{titlepage}

\pagenumbering{roman}

\begin{center}
    \hspace{0pt}
    \vfill
    
    \textit{Sometimes it is the people no one} \\
    \textit{imagines anything of who do the} \\
    \textit{things that no one can imagine.}
    
    \vspace{5mm}
    \textit{--- Alan Turing ---}
    
    \vfill
    \hspace{0pt}
    \pagebreak
\end{center}
\chapter*{Abstract}
\addcontentsline{toc}{chapter}{Abstract}

Developing increasingly efficient and accurate algorithms
for \emph{approximate nearest neighbor search} is a paramount
goal in modern information retrieval.
A primary approach to addressing this question is 
\emph{clustering}, which involves partitioning the dataset
into distinct groups, with each group characterized by a 
representative data point.
By this method, retrieving the top-$k$ data points for a query 
requires identifying the most relevant clusters based on their 
representatives---a \emph{routing} step---and then conducting 
a nearest neighbor search within these clusters only, 
drastically reducing the search space.

The objective of this thesis is not only to provide a 
comprehensive explanation of 
\emph{clustering-based approximate nearest neighbor search} 
but also to introduce and delve into every aspect of our novel 
state-of-the-art method, which originated from a natural 
observation: The routing function solves a ranking problem, 
making the function amenable to \emph{learning-to-rank}.
The development of this intuition and applying it to 
\emph{maximum inner product search} has led us to demonstrate 
that learning cluster representatives using a simple linear 
function significantly boosts the accuracy of clustering-based 
approximate nearest neighbor search.

\vfill

\subsubsection{Keywords} Information Retrieval; Clustering; Approximate Nearest Neighbor Search; Learning to Rank.
\chapter*{Notation}
\addcontentsline{toc}{chapter}{Notation}

This section offers a concise reference 
detailing  the notation and special symbols
used throughout this thesis. \\

\setlength{\tabcolsep}{25pt}
\renewcommand{\arraystretch}{1.5}
\begin{longtable}{c m{9cm}}
    $a, b, c, \dots$ & Vector \\
    $q$ & Query point \\
    $u^\ast$ & Resulting element of a problem\\ 
    $A, B, C, \dots$ & Matrix \\
    $\mathcal{U}$ & Dataset \\
    $\mathcal{X}$ & Collection of vectors \\
    $\mathcal{D}$ & Collection of documents \\
    $\mathcal{I}$ & Collection of items \\
    $\mathcal{Q}$ & Collection of queries \\
    $\mathcal{S}$ & Exact set of the $k$ most similar vectors to the query \\
    $\tilde{\mathcal{S}}$ & Set of the top-$k$ most similar vectors to the query, returned by an ANN algorithm \\
    $\{x_i\}_{i=1}^N$ & Set of $N$ elements \\
    $\{\mu_i\}_{i=1}^L$ & Set of $L$ standard representative points \\
    $\{\nu_i\}_{i=1}^L$ & Set of $L$ learnt representative points \\
    $\boldsymbol{x}$, $(x_i)_{i=1}^N$ & Tuple of $N$ elements \\
    $\mathbb{Z}^+$ & Set of positive integers \\
    $\mathbb{R}^n$ & Euclidean $n$-dimensional vector space \\
    $\mathcal{C}(\cdot)$ & Clustering function \\
    $L$ & Number of clusters \\
    $\{c_i\}_{i=1}^L$ & Set of L non-overlapping clusters \\
    $\delta(\cdot, \cdot)$ & Distance function \\
    $\langle \cdot, \cdot \rangle$ & Inner product of two vectors \\
    $\lVert \cdot  \rVert_{i}$ & L$i$ norm of a vector \\
    $|\cdot|$ & Absolute value \\ 
    $|\cdot|$ & Cardinality of a finite set \\
    $\left[0, 1\right]$ & Unit interval \\  
    $f(\cdot, \cdot)$ & Ranking function \\
    $\tau(\cdot)$ & Routing function \\
    $\acute{\tau}(\cdot)$ & Learnt linear routing function \\
    $\tilde{\tau}(\cdot)$ & Learnt nonlinear routing function \\
    $\ell$ & Number of top clusters retrieved by a routing function \\
    $\varphi(\cdot)$ & Encoder (embedding model) \\
    $\eta(\cdot)$ & Numerical representation function \\
    $l(\cdot, \cdot)$, $\mathcal{L}(\cdot, \cdot)$ & Loss function \\
    $m(\cdot)$ & Evaluation metric \\
\end{longtable}


\chapter*{Code Implementation}
\addcontentsline{toc}{chapter}{Code Implementation}

To ensure transparency and reproducibility, 
the code underlying this thesis is publicly
available at \url{https://github.com/tomvek/mips-learnt-ivf}.

This repository contains the official
code implementation at the base of
our paper ``A Learning-to-Rank 
Formulation of Clustering-Based Approximate
Nearest Neighbor Search'' 
\cite{Vecchiato_2024_paper}, thereby 
providing the specific implementation 
of our new proposed method.
By utilizing this code, readers can 
replicate the experiments presented 
in this work, employ the code for 
future research, or integrate it 
into other systems or projects.

Specifically, the code implementation was
exclusively carried out in 
Python\footnote{\url{https://www.python.org}}.
The Python version and all required libraries are 
documented in the `README.md' of the aforementioned 
repository. The experiments were conducted on a 
server running Ubuntu\footnote{\url{https://ubuntu.com}}
$18.04.6$ LTS equipped with an
Intel\footnote{\url{https://www.intel.com}} 
Xeon Platinum $8276$L CPU, $503$GiB of RAM,
and an NVIDIA\footnote{\url{https://www.nvidia.com}} 
Tesla T$4$ $16$GiB GPU.

\tableofcontents
\newpage

\pagenumbering{arabic}

\pagestyle{fancy}
\fancyhead{}
\fancyhead[CE]{\small \textsc{Learning Cluster Representatives for Approximate Nearest Neighbor Search}}
\fancyhead[CO]{\small \textsc{\nouppercase{\leftmark}}}
\renewcommand{\headrulewidth}{0pt}
\setlength{\footskip}{1.5cm}

\chapter{Introduction}

It may be opportune to begin with the definition of Information Retrieval:
\begin{displayquote}
    Information Retrieval (IR) is the scientific discipline concerned with 
    the efficient and effective retrieval of relevant material from large
    collections to satisfy the information needs.
\end{displayquote}

\noindent IR is having a profound and far-reaching impact, shaping 
the way we live, work, and learn. This is evident in every facet of 
contemporary society, as demonstrated by our use of search engines,
recommendation systems, digital libraries, social media, and dialogue 
or question-answering systems among others. Extending its influence 
to significant fields such as education, research, healthcare, 
and business.

Taking an abstract standpoint, we can view IR as a gateway to knowledge.
And, as the English philosopher Francis Bacon famously stated in $1597$, 
\emph{`knowledge is power'}.

This ``game of knowledge'' in IR is fundamentally defined by two major 
players: a collection of documents, representing the entirely of 
information available within a system; and the user's query, which
describes the specific information the user seeks. The final objective 
is to return the best documents that most effectively address the query.
In the IR community this ``game'' is called \emph{top-$k$ retrieval}.

\begin{figure}
    \centering
    \includegraphics[width=1\linewidth]{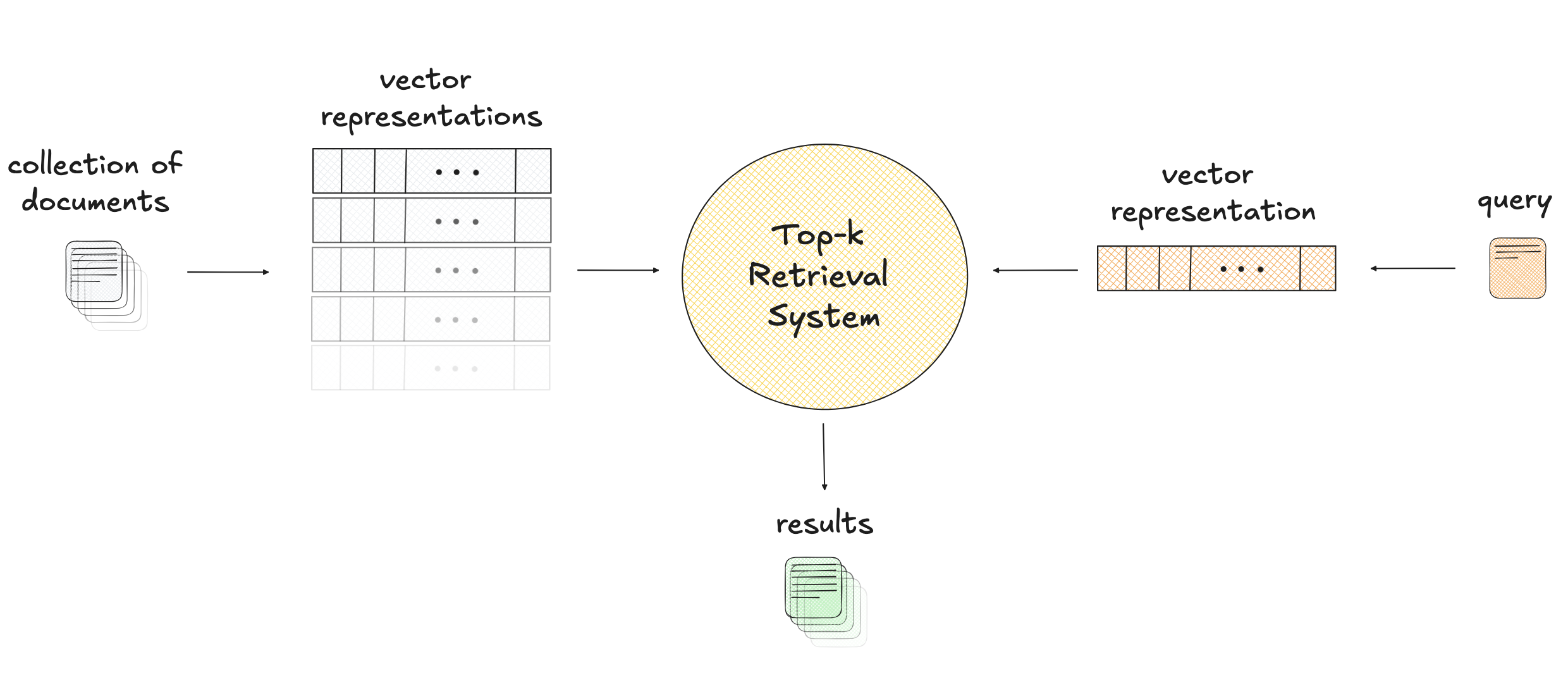}
    \caption{\footnotesize
    Top-$k$ retrieval problem based on the representation of documents
    and queries through vectors.
    }
    \label{fig:sketch-game-knowledge}
\end{figure}

One of the most prevalent and widely used methods for the top-$k$ retrieval
problem is to map both documents and queries into a vector space. 
A well-defined distance function is then applied to calculate the similarity
between  the corresponding vector representations, enabling the system to 
determine how well a document matches the user's query.
A visual representation is shown in Figure~\ref{fig:sketch-game-knowledge}.

However, despite its simplicity and elegance, this methodology 
presents a significant challenge: how can we optimally represent 
our documents and queries as vectors, ensuring that the actual 
similarity between these entities is accurately reflected in the
vector space?

The literature abounds with numerous works on the presented technique
and to provide increasingly accurate answers to the aforementioned
question. 
Among the earliest achievements, we find the vector space model based 
on the bag-of-words assumption~\cite{Salton_1962_Generation_paper, 
Salton_1975_Vector_paper}. Subsequently, other noteworthy methods 
emerged, such as TF-IDF~\cite{Salton_1988_TFIDF_paper} and BM$25$
~\cite{Robertson_1994_BM25_paper}. 
However, a significant advancement was achieved only when Machine 
Learning (ML)~\cite{Bishop_2006_ML_book} and Deep Learning (DL)
~\cite{Goodfellow_2016_DL_book} techniques were used to obtain the 
vector representation
~\cite{Huang_2013_DSSM_paper, Shen_2014_DSSMv2_paper, 
Zamani_2018_SNRM_paper, Lin_2021_PT_Transformer_book}. 
In particular, the turning point came with the Pre-trained Language 
Models (PLMs), which, based on Transformer architectures
~\cite{Vaswani_2023_Attention_paper}, have a deep understanding of the 
language and consequently representing documents and queries in an 
optimal manner: the resulting vectors are referred to as 
\emph{dense vectors} or \emph{embeddings}.
The remarkable success of PLMs, thanks especially to the 
\emph{pre-training followed by fine-tuning} paradigm, has led to 
the creation of many retrieval models based on them, opening the door
to \emph{(PLM-based) dense retrieval models}, a new generation of 
effective retrieval methods~\cite{Zhao_2022_Dense_Text_Retrieval_paper}.

With the dense vector representation in hand, the subsequent step involves 
retrieving the top-$k$ most relevant documents. To accomplish this, 
\emph{Nearest Neighbor (NN) search algorithms} are applied. NN algorithms 
perform an exhaustive search, computing the distance function 
between the query vector and each individual document vector within 
our collection.

But NN search algorithms suffer from a significant computational 
cost issue. In fact, for extensive document collections, applying
these algorithms during online search becomes infeasible. This poses 
another fundamental challenge given the ever-increasing volume of 
information and dataset sizes.
\emph{Approximate Nearest Neighbor (ANN) search}~\cite{Bruch_2024_book} 
has emerged as one of the most prevalent approach to tackle this 
problem and is currently a hot research topic in modern IR.

ANN search algorithms are a form of approximate retrieval and
are designed to trade off accuracy for speed. 
This implies that the output of the algorithm is often an 
approximation of the exact solution, but results are achieved 
significantly faster. In order to accomplish this goal, ANN 
search algorithms rely on data structures known as 
\emph{indexes}, which are constructed in an offline phase by 
processing the collection of vectors. Indexes are essential 
for space partitioning, enabling efficient navigation to 
locate the closest vectors to the query point, according to a 
distance function.
Conceptually, indexes act as guides that direct the search to 
specific regions of the data space, without examining the entire
dataset. Analogously, if one were searching for the Eiffel Tower, 
an index would guide the search directly to Paris, bypassing 
the need to search of all of France.

A prominent ANN search method is clustering-based approximation, 
where data points are grouped into $L$ geometric partitions using a 
clustering algorithm, and each partition is characterized by 
a representative vector. Partitions and their corresponding 
representative vectors constitute the index.
At search time, in the online phase, a distance function is computed
between the input query point and the representative vectors of the
clusters to determine the $\ell$ closest partitions, a process 
referred to as \emph{routing}. 
Once these partitions are identified, multiple strategies can be
adopted to extract the top-$k$ documents, such as employing an
another ANN algorithm or performing an exhaustive search with
these partitions.

In the standard clustering-based ANN search, the design 
choice include: employing Standard KMeans~\cite{Lloyd_1982_KMeans_paper} 
as the clustering algorithm to partition the data points into 
$L$ clusters; utilizing the mean of the cluster data 
points as the representative point; selecting 
the inner product (used in this work), $L1$ norm, 
$L2$ norm, or cosine similarity as the distance function; 
and setting $\ell$ significantly smaller than $L$ to 
substantially reduce the search space. At query time, 
to retrieve the top-$\ell$ partitions, the routing 
process is performed. Using the inner product as 
distance function, this process can be described by the
function $\tau(q) = \argmax^{\ell} Mq$, where 
$q \in \mathbb{R}^n$ is the query and 
$M \in \mathbb{R}^{L \times n}$ is a matrix whose rows 
correspond to the cluster centroids.

It is evident that the representative point is crucial in the routing process, as 
it determines the selection of partitions. Thus, this point must 
effectively encapsulate its cluster to maximize the accuracy of 
retrieving top-$k$ documents relative to the query within the 
returned partitions. Consequently, the following inquiries present 
themselves: Is the use of the mean the optimal choice?
How can we obtain the vector that best represents the partition?

In this work, we address these questions, developing a novel 
state-of-the-art clustering-based ANN search methodology 
and demonstrating that learning cluster representatives using 
a simple linear function significantly improves ANN search 
accuracy.

At the core of our research, we made a simple yet insightful 
observation:

\begin{displayquote}
    The \emph{routing function} solves a ranking problem, 
    making the function amenable to \emph{Learning-to-Rank}
    (LTR)~\cite{Bruch_2023_FTIR_book}.
\end{displayquote}

\noindent By developing this insight, we simply learn a linear function
$\acute{\tau}(q;\; W) = Wq$, where given a query $q \in \mathbb{R}^n$ 
as input and the learnt matrix $W \in \mathbb{R}^{L \times n}$, 
it returns the ranking scores for each of the $L$ partitions. 
Each row of the matrix $W$ corresponds to the 
\emph{learnt representative vector} for each cluster. 
Considering the top-$1$ scenario as an example, the function 
$\acute{\tau}$ ranks the partitions according to their likelihood of 
containing the nearest neighbor to the query.

Interestingly, all the required elements to learn this routing function are 
easily accessible: The training data consists of a set of queries; The 
ground-truth for each query comprises the partitions containing the exact 
top-$k$ documents; To determine the routing function's quality, we employ 
ranking metrics such as Mean Reciprocal Rank (MRR)~\cite{Liu_2009_LTR_IR_book}, 
which is appropriate for our task; Cross-entropy as loss function to maximize 
MRR~\cite{Bruch_2019_CrossEntropy_paper, Bruch_2021_CrossEntropy_paper} for 
the top-$1$ case, where each query has a single correct partition, and its 
generalization~\cite{Bruch_2021_CrossEntropy_paper} for top-$k$ with $k > 1$.

Through experiments on diverse text datasets, embeddings and 
clustering algorithms, we demonstrate empirically that learning
a simple linear routing function leads to significant accuracy 
gains, establishing a new advancement in clustering-based 
ANN search.

\section{Why This Thesis Matters}

\emph{Approximate Nearest Neighbor (ANN) search} is a hot
topic today that implicitly influences our lives.
Its pervasive use touches upon areas such as recommendation 
systems, image and video search, natural language processing,
fraud detection, music streaming services, and many more.
ANN search is a focal point of interest within the realm 
of vector databases. Numerous companies utilize ANN search 
algorithms and invest in them. Several prominent models,
such as ColBERTv$2$~\cite{Santhanam_2022_ColBERTv2_paper} 
and PLAID~\cite{Santhanam_2022_PLAID_paper}, incorporate 
this technique. Moreover, it is a widely studied research 
topic within the IR community due to its substantial impact.

\emph{Clustering-based ANN search} is one of the primary methods 
for ANN search and is central to the aforementioned discussion.
This thesis not only provides an in-depth explanation of this
methodology, delving into its intricacies and foundational 
principles to ensure a comprehensive understanding, but also 
introduces a novel state-of-the-art for clustering-based ANN 
search through a method we have developed~\cite{Vecchiato_2024_paper}.
Additionally, our proposed method is distinguished by its ease 
of integration into existing production systems utilizing 
clustering-based ANN search. By merely replacing the old 
centroids with the newly learnt representatives, the overall 
implementation remains unchanged, while significant accuracy 
improvements are realized.

Our findings demonstrate the potential of merging two major
IR fields: LTR and ANN. Motivating the community to explore 
this junction in future research.

In summary, we make the following contributions in this work:
\begin{itemize}
    \item We provide an in-depth explanation of clustering-based
          ANN search and present a general overview of LTR,
          outlining its core concepts;
    \item We present our novel clustering-based ANN search 
          algorithm that is based on learning cluster
          representatives;
    \item We show, by means of extensive experiments, 
          significant gains in accuracy using our proposed 
          method compared to the state-of-the-art baseline,
          for both top-$1$ and top-$k$ cases;
    \item We give a detailed analysis of our method, along
          with alternatives and variations.
\end{itemize}

\section{Organization}

The thesis is structured into six chapters.

Chapter $2$, $3$ and $4$ respectively delve into Vector
Search, Clustering-Based Approximate Nearest Neighbor 
Search and Learning-to-Rank. 
The following chapters are essential for a complete 
understanding of the research presented in the Chapter $5$. 
In particular, Chapter $2$ introduces Vector Search, 
starting from the concept of a vector and focusing on 
dense vectors. The chapter concludes by presenting the 
Maximum Inner Product Search problem. 
Chapter $3$ discusses Clustering-Based Approximate Nearest
Neighbor Search, where the focus is on its two main
components: clustering and approximate nearest neighbor 
search, which are explained. Regarding clustering, 
we concentrate on centroid-based clustering algorithms, 
specifically Standard Kmeans, Spherical Kmeans and 
Shallow Kmeans.
Chapter $4$ addresses Learning-to-Rank, providing an
overview of the topic and explaining what it means 
to learn a ranking function and the necessary ingredients 
for doing so.

Chapter $5$ provides our research project in detail.
We start by offering a general overview and then delve 
into the technical methodology, ensuring a comprehensive 
understanding of learning cluster representatives for 
Approximate Nearest Neighbor search, building upon the 
concepts introduced in previous chapters. Subsequently, 
the experimental setup and results are presented, 
demonstrating empirically that our proposed approach 
outperforms the state-of-the-art baseline. The chapter 
concludes with further analysis of our method and 
by exploring variations of it.

The general strategy employed in Chapters $2$, $3$, $4$ 
and $5$ involves first presenting the intuition and 
the idea underlying the method under consideration, 
followed by a concrete explanation.

Finally, Chapter $6$ conclude the thesis by recapitulating 
the main points and outlining potential avenues for 
future work.

The front and the end matter of the thesis contains: 
Abstract, Acknowledgements, Notation, Code Implementation,
and Bibliography.

\chapter{Vector Search}

\emph{Vector search} is a retrieval technique 
based on finding similar items, represented 
as vectors, within large collections by 
considering their semantic and contextual
meanings, all while maintaining efficiency.
In essence, vector search transforms the 
vast and chaotic expanse of data into a 
navigable and manageable world through the
utilization of vectors.

This technique leverages Machine Learning (ML)
models, commonly referred to as 
encoders~\cite{Xia_2020_BERT_survey_paper, 
Radford_2021_CLIP_paper}, 
to transform data into high-dimensional vectors, 
also know as dense vectors or embeddings. 
As the term \emph{embedding} implicitly 
suggests, these vectors encapsulate the 
semantic meaning behind the data they represent
within their numerical representation.
It is worth noting that the nature of the 
data we are discussing can be of various 
types, including text, images, audio, 
tabular data, and others.
For the sake of simplicity, we will 
exclusively focus on textual data in the 
remainder of this document.

In this context, we can conceptualize an 
encoder model as a function $\varphi$
that maps a document $x \in \mathcal{D}$, 
where $\mathcal{D}$ is the collection of 
documents, to a vector $d$: 
$\varphi(x) \rightarrow d \in \mathbb{R}^n$.

A particularly effective strategy within 
this domain is to think of vectors as 
points in a multidimensional space, with 
each vector element defining a position 
along a particular dimension.

The fundamental property of the 
$\varphi$ function, which is 
exploited by this technique, is that the 
semantic and contextual similarity between
items is reflected in their vector 
representation. Consequently, conceptually 
similar items will be located in close 
proximity within the high-dimensional space.
This implies that we can employ a distance 
function $\delta$ between these vectors as 
a measure of similarity to determine which 
items are semantically related. 

\begin{figure}
    \centering
    \includegraphics[width=1\linewidth]{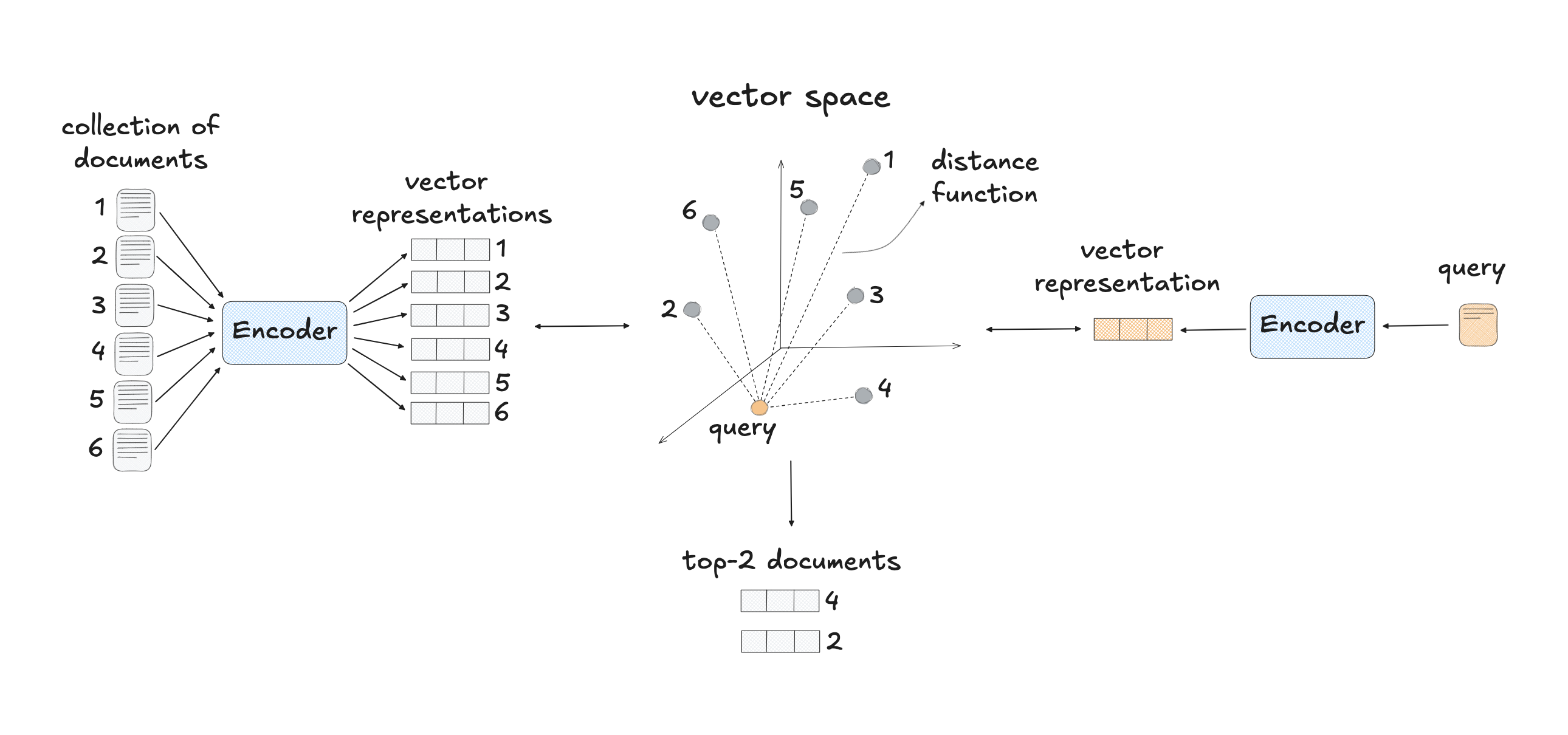}
    \caption{\footnotesize
    A visual representation of the vector search problem 
    within the $\mathbb{R}^3$ space. Given a collection of 
    documents, they are transformed into vector representations 
    using an encoder. During online search, a query $q$ is 
    vectorized using the same encoder, and a distance function 
    is applied to retrieve the top-$k$ similar documents.
    In practical applications, not only are the number of documents
    extremely large, but the vectors themselves are often 
    represented by hundreds or thousands of dimensions.
    }
    \label{fig:vector-search-idea}
\end{figure}

Given a query point $q \in \mathbb{R}^n$ and 
a collection of data points 
$\mathcal{X} \subset \mathbb{R}^n$ derived 
from $\mathcal{D}$ using $\varphi$, we
can compute the distance function between $q$ 
and each individual data point 
$u \in \mathcal{X}$ and, based on the score of
the distance metric, retrieve the top-$k$ most
semantically similar documents, which will be in
the nearest neighborhood of our query.
Formally, we can write:
\begin{equation}
    \label{equation:vector-search}
    \argmin_{u \in \mathcal{X}}^{(k)} \delta(q, u).
\end{equation}
Since a lower distance value corresponds to 
a higher degree of similarity.
Figure~\ref{fig:vector-search-idea} provides 
a visual depiction of the concepts discussed 
thus far.

Having explored vector search and
its three core components---dense vectors,
encoders (embedding models), and 
distance functions---we will now delve 
deeper into each.
Before proceeding, it's crucial to note 
that the relationship between items in the 
high-dimensional space vary based on the 
chosen encoder, which may focus on certain 
semantic aspects over others, and the 
distance function employed.

\section{Dense Vector Retrieval}

The concept of representing text 
as vectors is not new; 
early forms of such representations
emerged between $1950$ and $2000$,
with techniques like 
bag-of-words~\cite{Salton_1962_Generation_paper, Salton_1975_Vector_paper}, 
TF-IDF~\cite{Salton_1988_TFIDF_paper}, 
PPMI~\cite{Church_Hanks_1989_PPMI_paper, Dagan_Etal_1993_PPMI_paper}, 
LSI~\cite{Deerwester_1990_LSI_paper, Rosario_2000_LSI_paper}, 
and BM$25$~\cite{Robertson_1994_BM25_paper}. 
Subsequent advancements were 
driven by Machine Learning (ML), 
initially leading to 
\emph{static embedding} through 
models as 
Word$2$Vec~\cite{Mikolov_2013_Word2Vec_paper, Mikolov_2013_Word2Vec_Negative_paper}, 
GloVe~\cite{Pennington_2014_GloVe_paper},
and FastText~\cite{Bojanowski_2017_FastText_paper}, 
and later with more sophisticated
and expressive \emph{dynamic 
(or contextual) embeddings}\footnote{While static embeddings provide a single, 
                                     global representation for a given word, 
                                     subword o $n$-gram, contextual embeddings 
                                     can assign also different representations 
                                     based on the surrounding context, leading 
                                     to a more accurate understanding of the 
                                     word's, subword's or $n$-gram's meaning.}. 
Among the first prominent models 
for obtaining dynamic embeddings
was ELMo~\cite{Peters_2018_ELMo_paper}. 
However, the real turning point
came in $2018$ with the 
introduction of 
Transformers~\cite{Vaswani_2023_Attention_paper}, 
whose architecture radically 
changed the landscape of 
dynamic embedding representation. 
Since then, associated 
technologies like 
BERT~\cite{Devlin_2019_BERT_paper}, 
GPT~\cite{Radford_2018_GPT_paper}, 
XLNet~\cite{Yang_2020_XLNet_paper}, 
and T5~\cite{Raffel_2023_T5_paper} 
have consistently enhanced semantic
vector representations, particularly
after $2020$ with the advent of 
\emph{``Large'' Transformers}.
Large Transformers are Transformers 
with a larger architecture and 
trained on substantially larger 
datasets (e.g., GPT-3~\cite{Brown_2020_GPT3_paper} 
or GPT-4~\cite{Openai_2024_GPT4_paper}).
These models have demonstrated an 
exceptional ability to understand 
the language and map text to 
high-dimensional vector spaces, 
capturing intricate semantic nuances.

While our focus lies on \emph{dense
vectors}, it is important to also
acknowledge the existence of 
\emph{sparse vectors} for 
text representation.
Sparse vectors are extremely 
high-dimensional, containing 
many zero values, and the 
information is sparsely located.
They are optimal for tasks 
that rely on syntax, lexicon,
and exact or fuzzy text matching 
and are generated by algorithms 
such as BM$25$, TF-IDF,
and SPLADE~\cite{Formal_2021_SPLADE_paper}.

However, sparse vectors, due to 
their sparse nature, are generally 
less computationally efficient 
than dense vectors.
Indeed, with the latter, we can 
operate on contiguous regions of 
memory, unlike the former.
In addition, dense vectors, compared 
to sparse vectors, generally capture 
semantic meaning and underlying abstract 
concepts of a text more effectively.
This makes them preferable for 
conducting efficient and high-performing 
semantic search.

While still high-dimensional, 
dense vectors have a 
significantly lower 
dimensionality than sparse 
vectors, containing mostly 
non-zero values, and 
effectively utilize the entire 
space to represent the text.
Vector search models are based 
on dense vectors. Dense vectors 
are typically generated by 
machine learning models such as 
Transformers.

The vector space where dense 
vectors, generated by a given 
model, reside has a well-defined  
structure: semantically and 
contextually similar objects are 
mapped close together, while 
dissimilar objects are mapped 
far apart. The concept of 
distance and proximity within 
this space, and thus the 
similarity or dissimilarity 
between elements, is precisely 
defined by a distance function 
$\delta$.
This structure is particularly 
well-suited for the fundamental 
problem of top-$k$ retrieval, 
where vectors serve as retrieval 
units and the relevance between 
a query and a document is determined 
by measuring their similarity.

At the heart of numerous applications, 
including web search, recommendation 
systems, question answering systems, 
legal search, chatbots, and more, lies 
the fundamental problem of top-$k$ 
retrieval~\cite{Bruch_2024_book}. 
This problem is central to 
vector search and involves finding and 
retrieving the $k$ most similar objects 
within a collection given a query $q$.
By representing objects as vectors and 
defining similarity using a distance 
function $\delta$, 
the top-$k$ retrieval problem 
reduces to finding the $k$ points 
that minimize $\delta$ 
with respect to the query.

Let us now formalize the 
aforementioned intuition in 
Definition~\ref{definition:top-k-def}.
\begin{definition}[Top-$k$ Retrieval]
    \label{definition:top-k-def}
    Let be given a query point $q \in \mathbb{R}^n$,
    a collection of data points 
    $\mathcal{X} \subset \mathbb{R}^n$ and a 
    distance function $\delta(\cdot, \cdot)$, 
    the data point $u^\ast \in \mathcal{X}$ is 
    one of the top-$k$ data points, if
    \begin{equation}
        \label{equation:one-top-k-retrieval}
        u^\ast \in \argmin_{u \in \mathcal{X}}^{(k)} \delta(q, u).
    \end{equation}
\end{definition}

Thus far, we have considered the
embedding function (encoder) 
$\varphi$ and the distance
function $\delta$ as given. 
In Section $2.2$ and $2.3$, we 
will explore their internal 
mechanisms.

\section{Embedding Models}

There exist various embedding models 
for creating dense vector representations,
some of which were mentioned in the 
previous section.
However, the predominant approach involves 
the use of Transformer-based
Pre-trained Language Models (PLMs), 
due to their superior performance~\cite{Craswell_2023_TREC2022_paper}.

PLMs are (Large) Language Models (LLMs)~\cite{Minaee_2024_LLMs_Survey_book} 
pre-trained on massive datasets in a 
self-supervised~\cite{Balestriero_2023_SSL_book} 
manner, enabling them to understand 
language and capture a wide range of 
syntactic and semantic properties 
across diverse linguistic contexts.
Moreover, these models can be fine-tuned 
for specific downstream tasks.
Fine-tuning involves further training the 
PLM on specific task-oriented datasets, 
leveraging the pre-acquired knowledge to
specialize it for a particular domain.
This paradigm of 
\emph{``pre-training followed by fine-tuning''}
has enable state-of-the-art performance 
across a broad spectrum of tasks~\cite{Peters_2018_ELMo_paper, 
Howard_2018_FineTuning_paper, 
Devlin_2019_BERT_paper, Raffel_2023_T5_paper}.

Transformer-based PLMs are built upon 
the Transformer architecture, an 
encoder-decoder neural network model,
based on the \emph{self-attention} 
mechanism~\cite{Vaswani_2023_Attention_paper}.
Self-attention is the cornerstone for 
enabling the model to understand the 
underlying meaning of language and 
construct an internal mathematical 
representation of it. Specifically, 
self-attention effectively captures
the relationships and dependencies 
inherent in data by attending to 
how the different components of the 
text influence each other.
Transformers are inherently highly 
parallelizable, enabling them to 
process large amounts of data 
simultaneously. This characteristic 
allows for efficient pre-training 
of large language models.
Additional characteristics include 
the ability to effectively handle 
long text sequences and scale very 
well with massive datasets.

Based on variations in the original
Transformer architecture, training 
methods, model dimensionality in 
terms of parameters, dataset type and 
size, number of encoder and/or decoder
layers, and other differences, a 
multitude of Transformer-based 
Pre-trained Language Models (PLMs) have 
emerged, including BERT~\cite{Devlin_2019_BERT_paper}, 
DistilBERT~\cite{Sanh_2020_DistilBERT_paper}
RoBERTa~\cite{Liu_2019_RoBERTa_paper}, 
ALBERT~\cite{Lan_2020_ALBERT_paper},
XLNet~\cite{Yang_2020_XLNet_paper},
LLaMA~\cite{Touvron_2023_LLaMA_paper},
GPT-1~\cite{Radford_2018_GPT_paper},
GPT-2~\cite{Radford_2019_GPT2_paper},
GPT-3~\cite{Brown_2020_GPT3_paper},
GPT-4~\cite{Openai_2024_GPT4_paper},
T5~\cite{Raffel_2023_T5_paper},
mT5~\cite{Xue_2021_mT5_paper},
BART~\cite{Lewis_2019_BART_paper},
and many others.

The key point of interest is that 
all Transformer-based Pre-trained 
Language Models, whether fine-tuned
or not, and regardless of their specific 
characteristics, can generate dense 
vector representations of input texts 
that encapsulate their semantic meaning.

This section has introduced modern 
embedding models and provided a basic 
understanding of the $\varphi$
function. A more in-depth discussion 
of the presented models and their 
specific operations is beyond the 
scope of this work.
Readers interested in a deeper dive 
into this topic may consult
~\cite{Minaee_2024_LLMs_Survey_book,
Fan_2022_PLMs_IR_book, 
Zhou_2023_history_PLMs_book,
Han_2021_history2_PLMs_book,
Minaee_2021_history_DL_book,
Qiu_2020_history_PLMs_NLP_paper,
Liu_2020_history_Context_Embeddings_paper,
Zhao_2022_Dense_Text_Retrieval_paper,
Lin_2021_PT_Transformer_book,
Xia_2020_BERT_survey_paper}.

\section{Distance Functions}

Once vector representations of both 
the corpus documents and the input 
query are obtained, the subsequent and 
necessary step to retrieve the top-$k$ 
documents is to compute the distance 
function between the query point and 
each document point in the collection 
(Definition~\ref{definition:top-k-def}).
We will now define the distance function,
reformulating the top-$k$ retrieval problem
according to the specific metric used.

\emph{Manhattan distance}, 
\emph{Euclidean distance}, 
\emph{Cosine distance}, and 
\emph{Inner Product distance} are
the most frequently employed metrics
for vector distance calculation.

\begin{figure}
    \centering
    \begin{subfigure}[b]{0.20\textwidth}
        \centering
        \includegraphics[width=\textwidth]{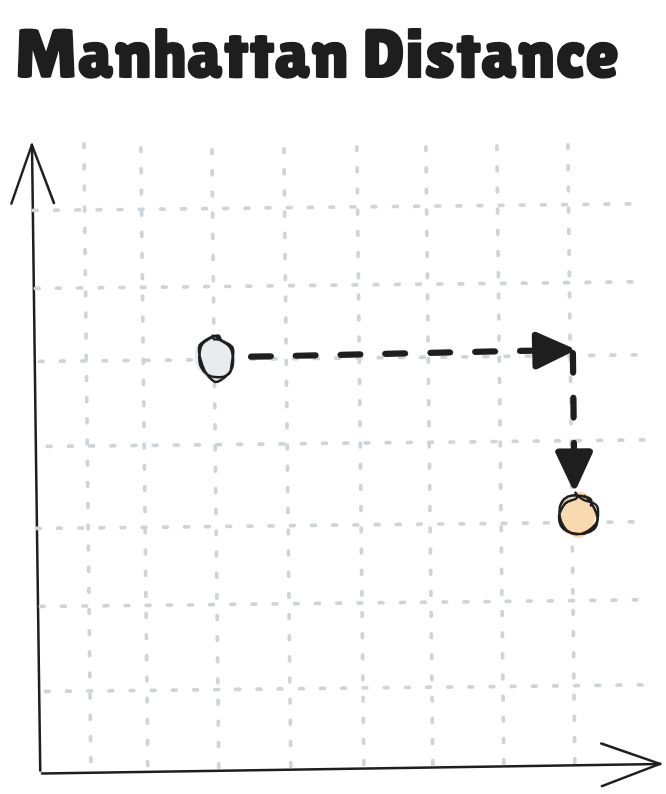}
        \caption{\footnotesize }
        \label{fig-a:manhattan-distance}
    \end{subfigure}
    \hfill
    \begin{subfigure}[b]{0.20\textwidth}
        \centering
        \includegraphics[width=\textwidth]{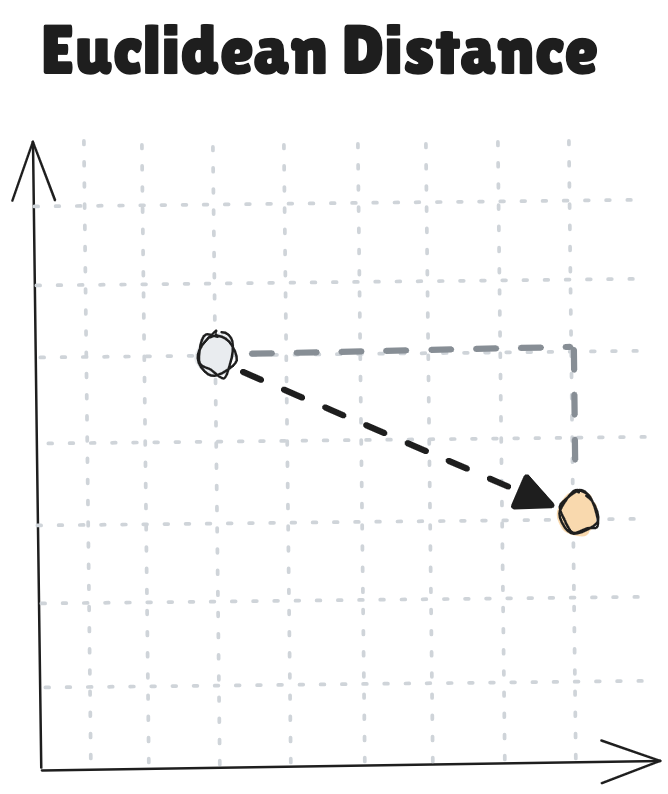}
        \caption{\footnotesize }
        \label{fig-b:euclidean-distance}
    \end{subfigure}
    \hfill
    \begin{subfigure}[b]{0.20\textwidth}
        \centering
        \includegraphics[width=\textwidth]{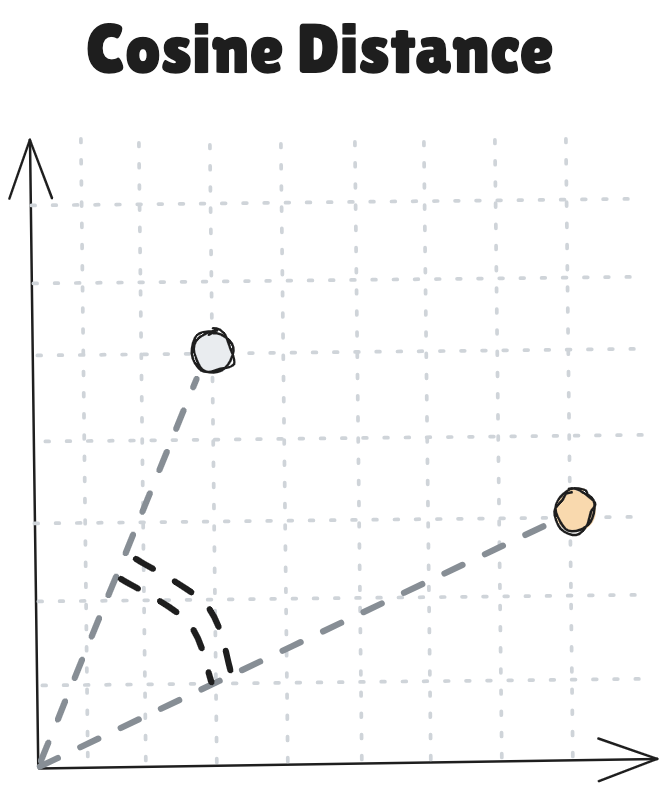}
        \caption{\footnotesize }
        \label{fig-c:cosine-distance}
    \end{subfigure}
    \hfill
    \begin{subfigure}[b]{0.21\textwidth}
        \centering
        \includegraphics[width=\textwidth]{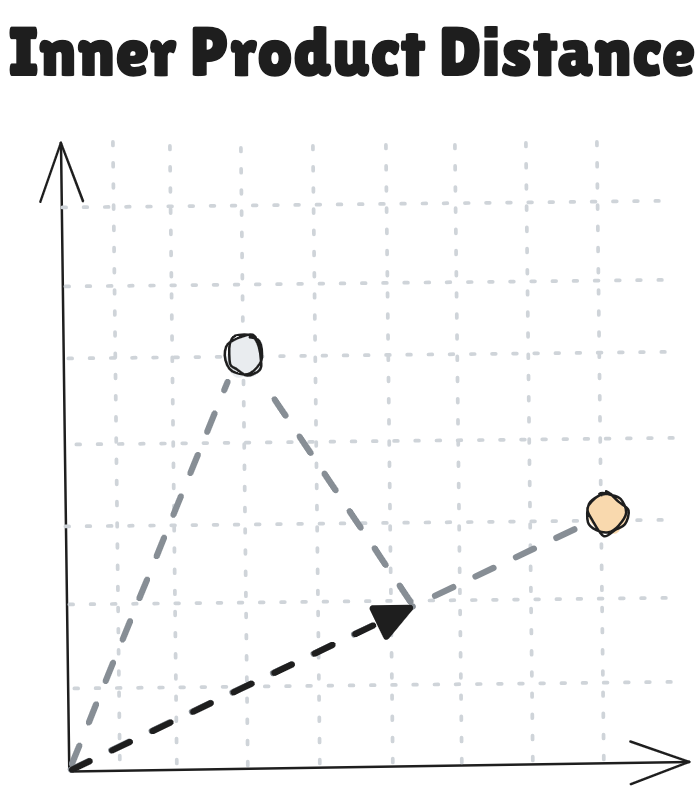}
        \caption{\footnotesize }
        \label{fig-d:inner-product-distance}
    \end{subfigure}
    \caption{\footnotesize
    Distance metrics applied to two 
    vectors, represented by the 
    gray and orange dots, in a 
    two-dimensional space
    $\mathbb{R}^2$.
    (a) Manhattan distance, L$1$ norm;
    (b) Euclidean distance, L$2$ norm;
    (c) Cosine distance;
    (d) Inner Product distance.
    }
    \label{fig:distance-metrics}
\end{figure}

\subsection{Manhattan Distance}

The Manhattan distance, or L$1$ norm, 
(Figure~\ref{fig-a:manhattan-distance})
measures the distance between two
points in a grid-like pattern, analogous 
to the distance traveled by a taxicab 
along city blocks in $\mathbb{R}^2$. 
It is computed by summing the absolute 
differences of the corresponding coordinates 
of the two points. In other words, imagine 
two points in an n-dimensional space
$\mathbb{R}^n$; the Manhattan distance 
from one point to the other is the sum 
of the distances traveled along each 
coordinate axis to reach the destination.

Formally, L$1$ norm between two points 
$q, u \in \mathbb{R}^n$ is defined as:
\begin{equation}
    \label{equation:manhattan-distance}
    \delta(q, u) = \lVert q - u  \rVert_{1} = \sum_{i=1}^{n} |q_i - u_i|
\end{equation}

The top-$k$ retrieval problem using 
the L$1$ norm as a distance function 
to compute similarity between the 
query and documents is referred to 
as $k$-Nearest Neighbors search with 
L$1$ norm ($k$-NN$_{1}$).
\begin{definition}[$k$-Nearest Neighbors search with L$1$ norm]
    \label{definition:k-NN-1}
    Let be given a query point 
    $q \in \mathbb{R}^n$ and a 
    collection of data points 
    $\mathcal{X} \subset \mathbb{R}^n$,
    the $k$-Nearest Neighbors search 
    with L$1$ norm ($k$-NN$_{1}$) 
    consists in finding:
    \begin{equation}
        \label{equation:k-NN-1}
        \{u^\ast_1, u^\ast_2, \cdots,  u^\ast_k\} \subseteq \argmin_{u \in \mathcal{X}}^{(k)} \lVert q - u  \rVert_{1}.
    \end{equation}
\end{definition}

\subsection{Euclidean Distance}

The Euclidean distance, also known as
the L$2$ norm,
(Figure~\ref{fig-b:euclidean-distance})
measures the distance between two
points in straight-line. 
It is computed by taking the square 
root of the sum of the squared 
differences of the corresponding 
coordinates of the two points under 
consideration.
In simple terms, the Euclidean 
distance is the length of the line 
segment connecting the two points.

Formally, L$2$ norm between two vectors 
$q, u \in \mathbb{R}^n$ is defined as:
\begin{equation}
    \label{equation:euclidean-distance}
    \delta(q, u) = \lVert q - u  \rVert_{2} = \sqrt{\sum_{i=1}^{n} (q_i - u_i)^2}
\end{equation}

The top-$k$ retrieval problem using 
the L$2$ norm as a distance function 
is known as $k$-Nearest Neighbors 
search with L$2$ norm ($k$-NN$_{2}$).
\begin{definition}[$k$-Nearest Neighbors search with L$2$ norm]
    \label{definition:k-NN-2}
    Let be given a query point 
    $q \in \mathbb{R}^n$ and a 
    collection of data points 
    $\mathcal{X} \subset \mathbb{R}^n$,
    the $k$-Nearest Neighbors search 
    with L$2$ norm ($k$-NN$_{2}$) 
    consists in finding:
    \begin{equation}
        \label{equation:k-NN-2}
        \{u^\ast_1, u^\ast_2, \cdots,  u^\ast_k\} \subseteq \argmin_{u \in \mathcal{X}}^{(k)} \lVert q - u  \rVert_{2} = 
        \argmin_{u \in \mathcal{X}}^{(k)} \lVert q - u  \rVert_{2}^{2}.
    \end{equation}
\end{definition}

\subsection{Cosine Distance}

The cosine distance
(Figure~\ref{fig-c:cosine-distance})
is defined as one minus the cosine 
similarity.
The cosine similarity measures 
the similarity between two vectors 
in $\mathbb{R}^n$ based on the 
cosine of the angle between them,
irrespective of their magnitude.
It is calculated by computing the 
dot product of the two vectors and 
dividing it by the product of their 
Euclidean norms. Intuitively, the 
smaller the angular distance between 
vectors, the greater their similarity.

Formally, the cosine distance between 
two vectors $q, u \in \mathbb{R}^n$,
with an angle $\theta$ between them, 
is defined as:
\begin{equation}
    \label{equation:cosine-distance}
    \delta(q, u) = 1 - cos(\theta) = 1 - \frac{\langle q, u \rangle}{\lVert q \rVert_{2} \lVert u \rVert_{2}} =
    1 - \frac{\sum_{i=1}^{n} q_i u_i}{\sqrt{\sum_{i=1}^{n} q_i^2} \sqrt{\sum_{i=1}^{n} u_i^2}}
\end{equation}

The top-$k$ retrieval problem using 
the cosine distance as a distance 
function is referred to as $k$-Maximum 
Cosine Similarity Search ($k$-MCS).
\begin{definition}[$k$-Maximum Cosine Similarity Search]
    \label{definition:k-MCS}
    Let be given a query point 
    $q \in \mathbb{R}^n$ and a 
    collection of data points 
    $\mathcal{X} \subset \mathbb{R}^n$,
    the $k$-Maximum Cosine Similarity 
    Search ($k$-MCS) consists in 
    finding:
    \begin{equation}
        \label{equation:k-MCS}
        \{u^\ast_1, u^\ast_2, \cdots,  u^\ast_k\} \subseteq 
        \argmin_{u \in \mathcal{X}}^{(k)} 1 - \frac{\langle q, u \rangle}{\lVert q \rVert_{2} \lVert u \rVert_{2}} = 
        \argmax_{u \in \mathcal{X}}^{(k)} \frac{\langle q, u \rangle}{\lVert q \rVert_{2} \lVert u \rVert_{2}}.
    \end{equation}
\end{definition}

\subsection{Inner Product Distance}

The inner product distance, also known as
the dot product distance,
(Figure~\ref{fig-d:inner-product-distance})
between two vectors is defined as 
the negative of their inner product.
The inner product measures the 
similarity between two vectors in 
$\mathbb{R}^n$, considering both 
their magnitude and direction. 
In simple terms, the longer two 
vectors are, i.e., the greater 
their magnitude, and the more
closely aligned they are, i.e., 
the smaller the angle between them,
the higher their similarity.

Geometrically, the inner product 
can be interpreted as the projection 
of one vector onto the line defined 
by the other.
The similarity between the two vectors 
is then computed by multiplying the 
length of this projection by the 
length of the vector that defines 
the line. The result is positive if 
the vectors point in the same direction 
and negative otherwise.

Formally, the inner product distance 
between two vectors $q, u \in \mathbb{R}^n$
is defined as:
\begin{equation}
    \label{equation:inner-product-distance}
    \delta(q, u) = - \langle q, u \rangle = - \sum_{i=1}^{n} q_i u_i
\end{equation}

$k$-Maximum Inner Product Search 
($k$-MIPS) is a variant of the 
top-$k$ retrieval problem where 
similarity between the query point 
and a data point is measured using
the inner product.

The $k$-NN$_{2}$ and $k$-MCS problems
are both particular cases of the broader 
$k$-MIPS problem~\cite{Bruch_2024_book}.
\begin{definition}[$k$-Maximum Inner Product Search]
    \label{definition:k-MIPS}
    Let be given a query point 
    $q \in \mathbb{R}^n$ and a 
    collection of data points 
    $\mathcal{X} \subset \mathbb{R}^n$,
    the $k$-Maximum Inner Product 
    Search ($k$-MIPS) consists in 
    finding:
    \begin{equation}
        \label{equation:k-MIPS}
        \{u^\ast_1, u^\ast_2, \cdots,  u^\ast_k\} \subseteq 
        \argmin_{u \in \mathcal{X}}^{(k)} - \langle q, u \rangle = 
        \argmax_{u \in \mathcal{X}}^{(k)} \langle q, u \rangle.
    \end{equation}
\end{definition}

This work focuses on 
the $k$-MIPS problem.
\chapter{Approximate Nearest Neighbor Search}

Having explored the most widely used 
vector similarity metrics and defined 
the variants of the top-$k$ retrieval 
problem ($k$-NN$_{1}$, $k$-NN$_{2}$, 
$k$-MCS, and $k$-MIPS) in relation to 
them, we now turn our attention to 
applying an algorithm to retrieve the
top-$k$ data points given a query point.
For simplicity and given that the 
distance function used in this work is
the inner product distance, we focus our
discussion on the $k$-MIPS problem.

In order to accurately retrieve the 
top-$k$ most relevant documents for 
a given query, thus providing an exact 
solution to the $k$-MIPS problem, an 
\emph{exhaustive search algorithm}, 
also known as a brute-force algorithm, 
is required.
This algorithm first computes the inner 
product between the query point and 
every data point within the collection, 
and then retrieves the $k$ data points 
with the highest similarity scores.

Although this approach is conceptually 
straightforward, the associated 
computational cost is substantial, 
rendering it impractical and 
highly inefficient at query time, 
especially in modern systems that handle 
extremely large numbers of 
high-dimensional vectors.
As evidence, we can consider the time 
complexity of computing the dot product 
between a query point and every data 
point in the collection, which is 
$\mathcal{O}(|\mathcal{X}|n)$, where 
$|\mathcal{X}|$ represents the number of 
data points and $n$ represents the 
dimensionality of the vector space. This 
makes it evident that exhaustive search 
algorithm is prohibitive for billions 
of vectors with thousands of dimensions.

\emph{Approximate Nearest Neighbor (ANN) search}
emerges as the technique to address this 
challenge, providing an efficient and 
scalable solution.
The efficiency and scalability of an 
ANN search method are achieved by 
sacrificing perfect accuracy and 
introducing a degree of approximation 
in the search results. Indeed, 
ANN is an approximate retrieval 
technique that returns an approximate 
top-$k$ set as output, rather than an 
exact one.

In order to assess the effectiveness 
of an approximate top-$k$ solution 
provided by an ANN search algorithm, 
and quantify the accuracy of the search,
we calculate the proportion of data 
points in the returned solution that 
are also present in the exact top-$k$
set. Formally, this can be expressed as:
$|\mathcal{S} \cap \tilde{\mathcal{S}}| / k$,
where $\mathcal{S}$ represents the set 
of relevant items (exact top-$k$) and 
$\tilde{\mathcal{S}}$ represents the 
set of retrieved items returned by the 
ANN algorithm.

ANN search algorithms are instrumental 
in modern vector search systems, 
offering an optimal trade-off between 
accuracy and efficiency. By sacrificing 
a small degree of precision, ANN can 
rapidly identify a promising subset of 
vectors from the vector collection that 
is likely to contain the desired results.
This subset is then subjected to more 
rigorous search methods, such as a 
secondary ANN algorithm with a stronger 
emphasis on accuracy or, for sufficiently 
small subsets, an exhaustive search.

In essence, vector search systems 
operate as pipelines, beginning with the 
entire corpus and primarily leveraging 
ANN algorithms to iteratively reduce the 
search space, where at each stage, the 
algorithm selects a smaller subset of 
vectors, ultimately returning the top-$k$
most relevant results. These results can 
subsequently be fed into more sophisticated 
ranking algorithms for further refinement.

This approach enables the development of 
extremely fast search systems capable of 
achieving nearly flawless accuracy.

Technically, an ANN setup first builds 
a data structure---the index---from the
given collection of vectors, in an 
offline phase. This index is then used 
to rapidly identify the most similar 
vectors to a new query point.

The landscape of ANN algorithms is vast, 
encompassing a wide array of approaches, 
including 
tree-based~\cite{Bruch_2024_book, Bentley_1975_KDTrees_paper, 
Fukunaga_1975_BranchBound_paper, Dasgupta_2015_RPTrees_paper, 
Beygelzimer_2006_CoverTrees_paper, Arora_2018_HDindex_paper}, 
hashing-based~\cite{Bruch_2024_book, Indyk_1998_LSH_paper, 
Huang_2015_QALSH_paper, Gao_2014_DSH_paper, 
Weiss_2008_SpectralHashing_paper}, 
graph-based~\cite{Bruch_2024_book, Malkov_2020_HNSW_paper, 
Fu_2018_NSG_paper, Morozov_2018_IP_NSW_paper, 
Jayaram_2019_DiskANN_paper},
and clustering-based~\cite{Bruch_2024_book, Jegou_2011_PQ_paper,
Bruch_2024_Bridging_Dense_Sparse_MIPS_paper, 
Chierichetti_2007_ClusterPruning_paper, Auvolat_2015_ClusteringANN_paper}
methods. This work concentrates on the 
clustering-based approach, also known 
as the Inverted File (IVF) method.

\section{Clustering-Based ANN Search}

Clustering-based ANN search
is a prominent ANN search method, 
demonstrating strong empirical 
performance~\cite{Auvolat_2015_ClusteringANN_paper} 
and widespread adoption in production 
systems\footnote{\url{https://turbopuffer.com/blog/turbopuffer}, \\
\indent $\,\,\,$ \url{https://www.pinecone.io/blog/serverless-architecture}}.
This methodology is grounded in a simple
yet acute intuition: Leveraging the 
intrinsic clustering behavior of data 
points within the search space. 
By exploiting these natural partitions, 
top-$k$ data points for a query can be 
retrieved by examining solely the 
most similar clusters, thus 
significantly reducing the search space 
while maintaining a high level of 
retrieval accuracy.

More specifically, the clustering-based 
ANN methodology first takes the given 
collection of $P$ data points 
$\mathcal{X} \subset \mathbb{R}^n$ and 
applies a \emph{clustering} function to 
them, mapping each point to a specific 
cluster, $\mathcal{C}: \mathbb{R}^n \rightarrow \{1, 2, \dots, L\}$,
thereby partitioning the vectors under 
consideration.
Subsequently, each partition is 
represented by a vector.
The resulting partitions and their 
respective representative vectors 
constitute the \emph{index}.

\begin{figure}
    \centering
    \includegraphics[width=1\linewidth]{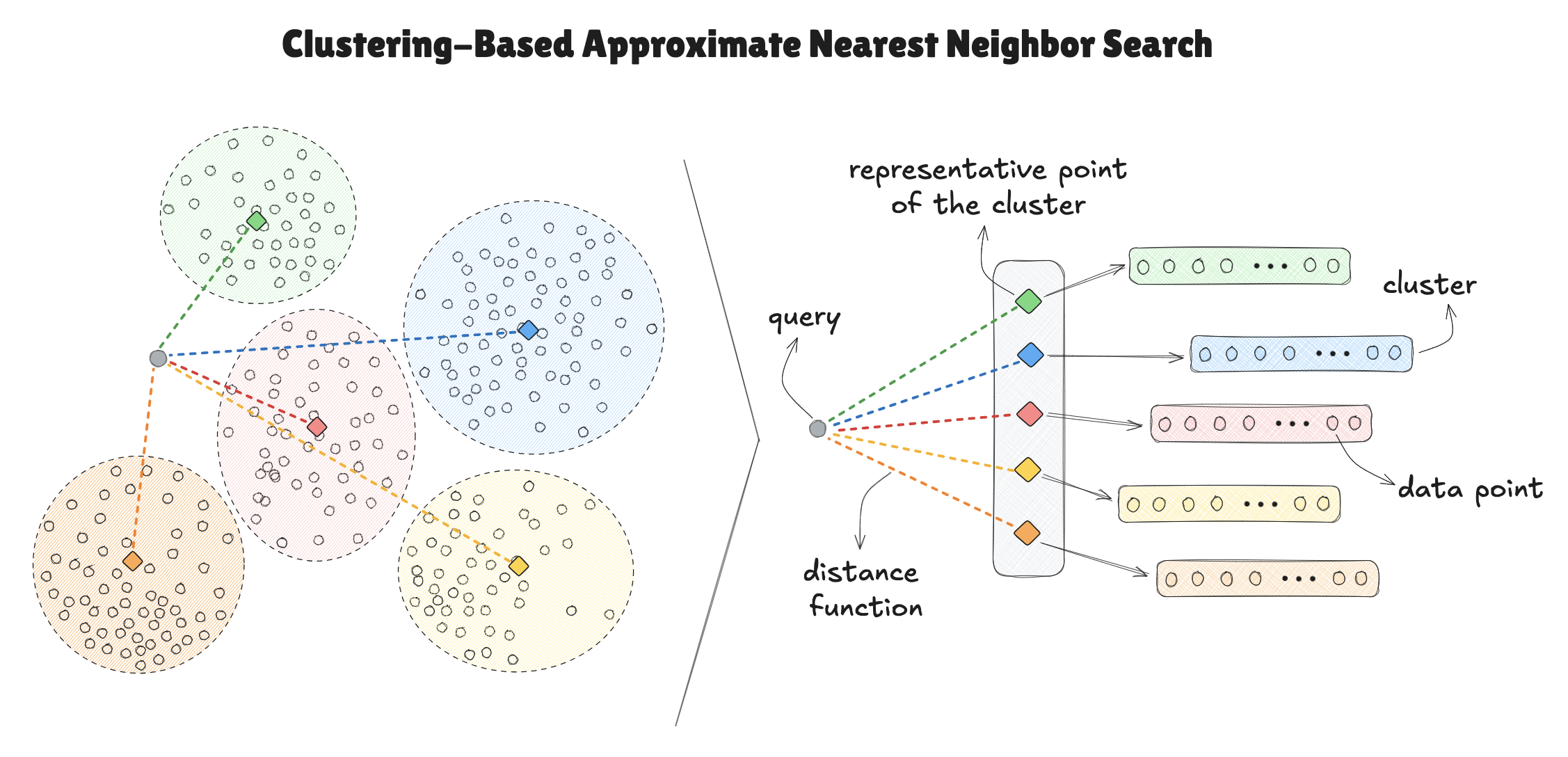}
    \caption{\footnotesize
    Clustering-Based Approximate Nearest 
    Neighbor (ANN) Search.
    The left-hand side presents a visual 
    representation of the space partitioned 
    into clusters, while the right-hand 
    side illustrates the corresponding index 
    structure. Both figures depict the 
    process of computing the similarity 
    between a query point and the 
    cluster representative points.
    }
    \label{fig:sketch-cluster-ann}
\end{figure}

At query time, when a query point $q$ is 
provided as input, a \emph{routing} function
$\tau: \mathbb{R}^n \rightarrow \argmax^{\ell}\mathbb{R}^L$
is invoked to calculate the similarity 
between the representative vectors and 
$q$. The function then returns the 
clusters associated with the top-$\ell$ 
most similar representative points to 
the query $q$.
Put differently, the routing function 
$\tau$ maps a query vector $q$ to the 
$\ell$  most similar partitions to it, 
within which the top-$k$ data points are 
most likely to be found. In order to 
retrieve the $\ell$ most similar 
partitions, $\tau$ essentially solves 
a top-$\ell$ retrieval problem where 
the collection of points is represented by 
the representatives points of clusters.
Figure~\ref{fig:sketch-cluster-ann} offers 
a visual depiction of the core ideas 
underlying the clustering-based ANN 
method discussed so far.

Once the $\ell$ partitions have been 
obtained, to retrieve the top-$k$ data 
points relative to the query $q$, 
a second search phase is conducted 
considering only the data points within 
those partitions. The search within 
these clusters can be exhaustive---as 
performed in this work---using a 
brute-force algorithm, or by employing 
another ANN algorithm.
It is trivial to observe that by using 
$\ell \ll L$, the search space for the 
second phase is significantly reduced, 
leading to more efficient search at the 
cost of potential accuracy loss.

A standard choice~\cite{Auvolat_2015_ClusteringANN_paper, 
Bruch_2024_book, Jegou_2011_PQ_paper} 
for clustering-based ANN is to use 
the Standard KMeans~\cite{Lloyd_1982_KMeans_paper} 
as the clustering algorithm $\mathcal{C}$, 
setting the number of clusters to
$L = \mathcal{O}(\sqrt{P})$, 
and defining the routing function $\tau$ 
as:

\begin{equation}
    \label{equation:tau-mips}
    \tau(q) =  \argmin_{i = 1, \dots, L}^{(\ell)} \delta(q, \mu_i),
\end{equation}

where $\mu_i$, the representative of the 
$i$-th cluster, is the cluster centroid 
defined as the mean of the data points 
within it:
$\mu_i = \frac{1}{|\mathcal{C}^{-1}(i)|}\sum_{u \in \mathcal{C}^{-1}(i)} u$,
with 
$\mathcal{C}^{-1}(i) \coloneq \{u \, | \, u \in \mathcal{X},\, \mathcal{C}(u) = i\}$.

As we have observed, clustering plays a 
pivotal role in this approach, with the
Standard KMeans algorithm serving as 
the default choice. Consequently, in the 
subsequent section, we delve deeper into 
these concepts. Firstly, we provide an 
overview of clustering, followed by a 
detailed explanation of the Standard 
KMeans and two other interesting 
variants employed in this work: 
Spherical KMeans~\cite{Dhillon_1999_Spherical_KMeans_report}
and Shallow KMeans~\cite{Chierichetti_2007_ClusterPruning_paper}.

\subsection{Clustering}

The initial question that must be 
addressed is: what is clustering?

\begin{displayquote}
    Given a collection of objects,
    \emph{clustering} is a technique 
    designed to group similar objects 
    into the same cluster and 
    dissimilar objects into different
    clusters.
\end{displayquote}

\noindent The application of clustering is 
ubiquitous, finding utility in fields 
such as machine learning, bioinformatics, 
pattern recognition, social network 
analysis, information retrieval, computer 
vision, climatology, healthcare, economics, 
and many more.
Its application across such diverse fields
is not surprising, given the inherent 
tendency of objects or data points to form 
natural clusters. This makes clustering 
an optimal approach for many applications.

It is crucial to emphasize that a 
clustering function, designed to uncover 
structures and patterns within a given 
collection of objects to create groups 
or partitions, necessitates a key 
element: a metric for comparing these 
objects. This metric is the core of 
the objective function. Indeed, 
clustering is an optimization problem 
where the goal is to achieve groups  
with the highest intra-cluster 
similarity and the lowest inter-cluster
similarity.

In relation to the problem described in 
the preceding chapters of retrieving the
top-$k$ documents with the highest 
accuracy and efficiency from a large 
collection of data points, it is intuitive 
that clustering aligns well 
with this task. As  discussed in 
Section 2.1, the vector representation 
has the property that semantically 
similar objects are mapped close together 
in the space, while dissimilar objects 
are mapped far apart. Hence, by clustering 
semantically similar objects, we can 
accelerate the search for the top-$k$ 
documents, because we can focus our 
attention solely on the vectors within
the clusters most similar to a given 
query $q$. Resulting in an effective 
ANN algorithm.

The literature presents a variety of 
types of clustering~\cite{Xu_2015_Survey_Clustering_paper}, 
including partition-based, density-based, 
distribution-based, hierarchical, and 
graph-based, each optimal in specific 
scenarios.
However, in this context, we will focus 
on the partitioning approach and 
examine the default algorithm employed
for clustering-based ANN, namely 
the KMeans algorithm.

\subsection{Standard KMeans}

Partition-based clustering is a type of 
clustering that groups data points into 
non-overlapping partitions. The KMeans 
algorithm is a prominent example of this 
type of clustering.

The KMeans algorithm is an iterative 
clustering algorithm that returns a 
local optimum for the KMeans optimization
problem, which is defined in~\ref{definition:k-means-problem}.

\begin{definition}[KMeans problem]
    \label{definition:k-means-problem}
    The KMeans problem, given $L \in \mathbb{Z}^+$ 
    and a collection of data points 
    $\mathcal{X} \subset \mathbb{R}^n$,
    consists of identifying $L$ centroids 
    $\{\mu_i\}_{i=1}^{L}$ that minimize the 
    objective function $\mathcal{E}$,
    defined as:
    \begin{equation}
        \label{equation:inertia}
        \mathcal{E}(\{\mu_i\}_{i=1}^{L}) = \sum_{x \in \mathcal{X}} \lVert x -  \argmin_{\mu_i \in \{\mu_i\}_{i=1}^{L}} \lVert x - \mu_i \rVert_2^2 \rVert_2^2.
    \end{equation}
\end{definition}

Considering the definition~\ref{definition:k-means-problem},
it is clear that problem is intrinsically 
linked to the concept of clustering. 
Specifically, the KMeans problem can be 
reformulated as finding the $L$ 
partitions of a collection of data points 
$\mathcal{X}$ such that the centroids of 
these partitions minimize the Equation~\ref{equation:inertia}.

The KMeans clustering algorithm aims to 
create $L$ partitions that minimize the 
objective function $\mathcal{E}$ 
(Equation~\ref{equation:inertia}), also 
known as inertia. To achieve this, after 
an initial centroids initialization phase, 
the algorithm iteratively performs two 
steps: first, each data point is assigned 
to the nearest centroid; then, the centroids 
are recalculated based on the new cluster 
memberships. This iterative process ensures 
a decrease in inertia until a local optimum 
is reached~\cite{David_2007_KMeanspp_paper}.
The algorithm terminates when a stopping 
criterion is satisfied, such as after 
$t$ iterations or when the change in 
$\mathcal{E}$ is less than a specified 
threshold, 
$|\mathcal{E}(\{\mu_i\}_{i=1}^{L})^{(i)} - \mathcal{E}(\{\mu_i\}_{i=1}^{L})^{(i+1)}| < \epsilon$.

The pseudocode for the Standard KMeans 
clustering algorithm is presented in 
Algorithm~\ref{alg:standard-kmeans}.

\begin{algorithm}
    \caption{Standard KMeans Algorithm}\label{alg:standard-kmeans}
    \begin{algorithmic}[1]
        \State \textbf{Input:} number of clusters $L$,  collection of data points $\mathcal{X} \subset \mathbb{R}^n$
        \State \textbf{Output:} cluster centroids $\left[\mu_1, \dots, \mu_L\right] \in \mathbb{R}^{n \times L}$, non-overlapping partitions $\{c_i\}_{i=1}^{L}$

        \State Randomly select $L$ cluster centroids $\mu_1, \dots, \mu_L$
        \Repeat
            \State  $\{c_i\}_{i=1}^{L} = \{\}$ 
            \For{\textbf{each} $x \in \mathcal{X}$}
                \State $c_i = c_i \cup \{x\}$, where $i \coloneq \argmin_{i=1, \dots, L} \lVert x -  \mu_i \rVert_2^2 $
            \EndFor
            \For{$i \in \{1, \dots, L\}$}
                \State $\mu_i = \frac{1}{|c_i|}\sum_{x \in c_i} x$
            \EndFor
        \Until {$\mu_1, \dots, \mu_L$ converge}
    \end{algorithmic}
\end{algorithm}

The KMeans algorithm is one of the most 
widely used, recognized, and popular 
clustering algorithms~\cite{Jain_2010_50Years_KMeans_paper,
Yin_2024_Rapid_View_Clustering_paper,
Xu_2015_Survey_Clustering_paper,
Ikotun_2023_KMeans_Review_paper,
Ahmed_2020_KMeans_paper,
Berkhin_2006_Survey_Clustering_paper}. 
In general, the Standard KMeans algorithm and 
its variants, in addition to being 
intrinsically simple to understand 
and implement, are also efficient algorithms
with low time complexity and high 
computational efficiency, making them 
well-suited to use on massive 
datasets~\cite{Xu_2015_Survey_Clustering_paper,
Yin_2024_Rapid_View_Clustering_paper,
Ikotun_2023_KMeans_Review_paper,
Capo_2018_Efficient_KMeans_paper,
Shindler_2011_Fast_KMeans_paper,
Xu_2005_Survey_Clustering_paper}.

However, the KMeans algorithm is not 
without its drawbacks, such as 
sensitivity to the initial choice of 
centroids, the requirement of specifying
the number of clusters a priori, 
difficulties in achieving a global minimum 
for complex datasets, sensitivity to noise 
data and outliers, and suboptimal 
performance on non-convex data; although 
numerous studies and variants have been 
proposed to address these issues
\cite{Xu_2015_Survey_Clustering_paper,
Ikotun_2023_KMeans_Review_paper,
Ahmed_2020_KMeans_paper,
Berkhin_2006_Survey_Clustering_paper,
David_2007_KMeanspp_paper}.

Let us now turn our attention to two 
variants of the Standard KMeans algorithm:
Spherical KMeans and Shallow KMeans.

\subsection{Spherical KMeans}
Both Spherical KMeans~\cite{Dhillon_1999_Spherical_KMeans_report} 
and Standard KMeans are commonly 
employed in clustering-based ANN~\cite{Bruch_2024_Bridging_Dense_Sparse_MIPS_paper}.

The fundamental distinction between the 
two aforementioned algorithms lies in 
their respective focus  on similarity 
measures and the consequent objective 
functions. While Standard KMeans 
algorithm minimizes an objective 
function based on Euclidean distance
(Definition~\ref{definition:k-means-problem}),
Spherical KMeans minimizes an objective 
function that employs cosine distance 
(Equation~\ref{equation:cosine-distance})
to form L partitions.
By L$2$-normalizing all $x \in \mathcal{X}$ 
to unit length, Spherical KMeans is an 
iterative algorithm that aims to find $L$ 
partitions maximizing the following objective 
function:
\begin{equation}
    \label{equation:spherical-kmeans-problem}
    \mathcal{P}(\{\mu_i\}_{i=1}^{L}) = \sum_{x \in \mathcal{X}} \langle x,  \argmax_{\mu_i \in \{\frac{\mu_i}{\lVert \mu_i \rVert_2}\}_{i=1}^{L}} \langle x, \mu_i \rangle \rangle.
\end{equation}

As with the KMeans problem~\cite{Aloise_2009_NPHard_Standard_paper}, 
determining the globally optimal 
partitions in this case is also 
NP-hard~\cite{Kleinberg_1998_NPHard_Spherical_paper}.
Thus, Spherical KMeans, like Standard 
Kmeans, is an approximation algorithm 
for the optimal solution.
Nonetheless, it is an efficient and 
effective iterative heuristic, prone 
to local optima, and capable of  
yielding reasonable results~\cite{Dhillon_1999_Spherical_KMeans_report}.

Algorithm~\ref{alg:spherical-kmeans} 
provides the pseudocode for the 
Spherical KMeans clustering algorithm.

\begin{algorithm}
    \caption{Spherical KMeans Algorithm}\label{alg:spherical-kmeans}
    \begin{algorithmic}[1]
        \State \textbf{Input:} number of clusters $L$,  collection of data points $\mathcal{X} \subset \mathbb{R}^n$
        \State \textbf{Output:} cluster centroids $\left[\mu_1, \dots, \mu_L\right] \in \mathbb{R}^{n \times L}$, non-overlapping partitions $\{c_i\}_{i=1}^{L}$

        \State Randomly select $L$ cluster centroids $\mu_1, \dots, \mu_L$ on the unit sphere
        \State $x = \frac{x}{\lVert x \rVert_2} \, \forall x \in \mathcal{X}$
        \Repeat
            \State  $\{c_i\}_{i=1}^{L} = \{\}$ 
            \For{\textbf{each} $x \in \mathcal{X}$}
                \State $c_i = c_i \cup \{x\}$, where $i \coloneq \argmax_{i=1, \dots, L} \langle x, \mu_i \rangle$
            \EndFor
            \For{$i \in \{1, \dots, L\}$}
                \State $\mu_i = \frac{1}{|c_i|}\sum_{x \in c_i} x$
                \State $\mu_i = \frac{\mu_i}{\lVert \mu_i \rVert_2}$ \Comment $\mu_i$ is projected on the unit sphere
            \EndFor
        \Until {$\mu_1, \dots, \mu_L$ converge}
    \end{algorithmic}
\end{algorithm}

Comparing Algorithm~\ref{alg:standard-kmeans} 
with Algorithm~\ref{alg:spherical-kmeans}, 
the Spherical KMeans algorithm introduces 
the following modifications: initial 
L$2$-normalization of all vectors $x \in \mathcal{X}$;
cluster assignment based on inner products 
between data points and centroids during 
iterations; and, the projection of 
centroids onto the unit sphere at the 
end of each iteration.

The Spherical KMeans clustering algorithm, 
as defined, is a suitable choice for 
clustering-based ANN with cosine distance 
(Equation~\ref{equation:cosine-distance})
or inner product distance 
(Equation~\ref{equation:inner-product-distance})
as the distance metric of interest.

\subsection{Shallow KMeans}
We now turn our attention to a 
particularly simple, computationally 
efficient, and interesting clustering 
algorithm, which we refer to as 
Shallow KMeans. The corresponding 
pseudocode is presented in 
Algorithm~\ref{alg:shallow-kmeans}.

\begin{algorithm}
    \caption{Shallow KMeans Algorithm}\label{alg:shallow-kmeans}
    \begin{algorithmic}[1]
        \State \textbf{Input:} number of clusters $L$,  collection of data points $\mathcal{X} \subset \mathbb{R}^n$
        \State \textbf{Output:} cluster representatives $\left[r_1, \dots, r_L\right] \in \mathbb{R}^{n \times L}$, non-overlapping partitions $\{c_i\}_{i=1}^{L}$

        \State Randomly select $L$ data points $s_1, \dots, s_L$ from $\mathcal{X}$
        \State  $\{c_i\}_{i=1}^{L} = \{\}$ 
        \For{\textbf{each} $x \in \mathcal{X}$}
            \State $c_i = c_i \cup \{x\}$, where $i \coloneq \argmin_{i=1, \dots, L} \delta(x, s_i) $
        \EndFor
        \State Select a representative point for each cluster $r_1, \dots, r_L$
    \end{algorithmic}
\end{algorithm}

Shallow KMeans can be viewed as a single 
iteration of the KMeans algorithm. 
It involves randomly selecting $L$ data 
points from $\mathcal{X}$ as initial 
cluster representatives, followed by  
assigning each data point to the nearest 
representative. The cluster 
representatives can then be updated, for 
example, by computing the mean of the 
points in each cluster or employing 
other strategies. However, this update 
step is not mandatory, and the initial 
$L$ points can be used as final 
representatives.
In our implementation, the distance 
function employed is the inner product 
distance (Equation~\ref{equation:inner-product-distance}),
and the final cluster representatives
coincide with the initial ones.

The Shallow KMeans clustering algorithm was 
originally introduced in~\cite{Chierichetti_2007_ClusterPruning_paper},
where the authors proposed an extremely 
simple pruning scheme, termed 
\emph{cluster pruning}, to efficiently 
address the top-$k$ retrieval problem 
(Definition~\ref{definition:k-means-problem})
while maintaining good retrieval quality.
The clustering algorithm, in particular, 
is used in the preprocessing phase of 
cluster pruning scheme with some variations.
Furthermore, \cite{Chierichetti_2007_ClusterPruning_paper} 
demonstrated that the proposed approach, 
and consequently the underlying 
clustering algorithm, is both efficient 
and accurate in retrieving the top-$k$ 
documents for a given query, achieving 
remarkable performance. As we will also 
see in this work.

\chapter{Learning-to-Rank}

To realize the proposed method, which 
will be detailed in the subsequent 
chapter, we have employed a widely 
studied and researched technique within 
the scientific community: 
\emph{Learning-to-Rank (LTR)}.
By developing a model that harnesses 
the power of LTR to enhance ANN search.
The results obtained highlight the 
potential of combining these two fields, 
thus motivating the community to 
further investigate into this intersection.

The objective of this chapter is to 
provide an overview of LTR and equip 
readers with the necessary knowledge 
to fully understand the ranking model 
employed in our work. For a deeper 
dive into LTR, we suggest referring 
to \cite{Liu_2009_LTR_IR_book, 
Bruch_2023_FTIR_book}.

We will now introduce
Learning-to-Rank through the 
following fundamental question: 
what is LTR?

\begin{displayquote}
    \emph{Learning-to-Rank (LTR)} is a 
    supervised Machine Learning (ML) 
    technique aimed at \emph{learning} a 
    function to solve a \emph{ranking}
    problem.
\end{displayquote}

\begin{figure}
    \centering
    \includegraphics[width=1\linewidth]{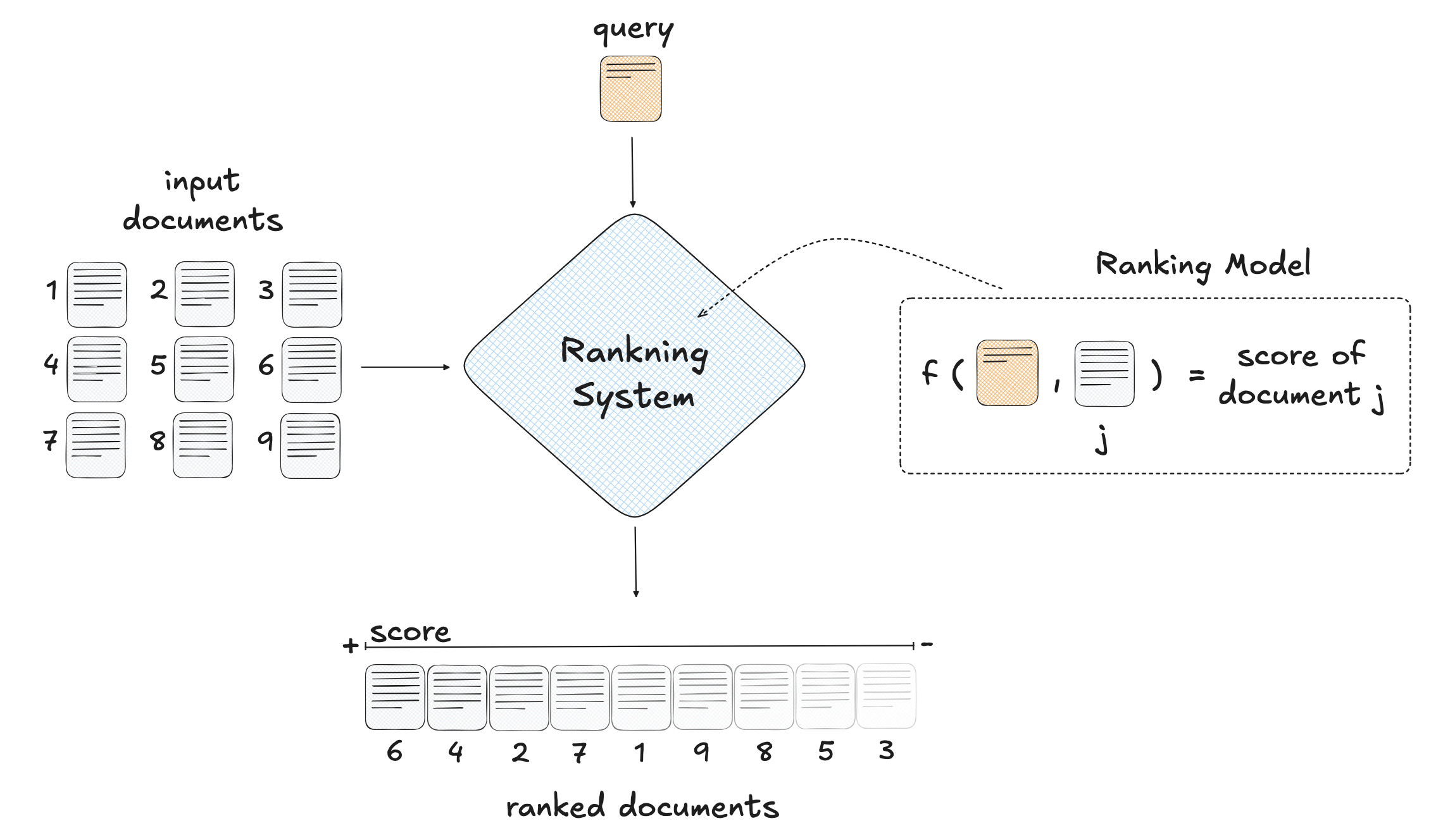}
    \caption{\footnotesize
    A ranking system is proposed that, 
    given a set of documents and a query 
    as input, outputs the documents 
    ordered by their relevance to the query. 
    This relevance is obtained through a 
    ranking model, which assigns a score 
    to each document-query pair, reflecting 
    the document's pertinence to the query.
    }
    \label{fig:sketch-ranking-model}
\end{figure}

\noindent More specifically, the learnt 
function is a ranking model or ranking 
function
$f: \mathcal{Q} \times \mathcal{I}^U \rightarrow \mathbb{R}^U$, 
which, given a query $q \in \mathcal{Q}$
and a set of $U$ items 
$\boldsymbol{d} = (d_i)_{i=1}^U \in \mathcal{I}^U$ 
as input, computes a score 
$s_i \in \mathbb{R}$ for each 
query-item pair: 
$f(q, (d_i)_{i=1}^U) = (s_i)_{i=1}^U = \boldsymbol{s} \in \mathbb{R}^U$. 
These scores are then used to order 
the items by relevance.
A visual depiction is provided in 
Figure~\ref{fig:sketch-ranking-model}.

To illustrate LTR more clearly, 
image our ranking function 
$f$ as a pastry chef.
The chef's task is to arrange 
a set of pastries according to a 
customer's preferences. By assigning 
a relevance score to each pastry 
based on the customer's order, the
chef create a ranked list. Through 
training and learning from mistakes, 
the chef can improve their ability 
and generate more accurate rankings.

Staying within the analogy, in the 
next section we will delve deeper 
into the pastry chef and explore how 
they actually learn.

\section{Learnt Ranking Function}

The learnt ranking function is the 
mathematical function learnt through 
supervised learning techniques by 
LTR models.
To learn this function, we need 
the following key ingredients.

\begin{itemize}
    \item \textbf{Dataset $\mathcal{U}$}
    
    The dataset $\mathcal{U}$ serves 
    as the raw material for learning 
    the function $f$, 
    providing the data required to train, 
    validate, and test it.
    Typically, an LTR dataset consists 
    of a set of triplets 
    $(q, \boldsymbol{d}, \boldsymbol{y}) \in \mathcal{Q} \times \mathcal{I}^U \times \mathcal{Y}^U$,
    where $q$ symbolizes the query, 
    $\boldsymbol{d} = (d_i)_{i=1}^U$  
    is the collection of items to be 
    ranked in relation to $q$, and 
    $\boldsymbol{y} = (y_i)_{i=1}^U$ 
    represents the ground-truth 
    relevance judgment 
    for each query-item pair.
    
    Formally,
    $\mathcal{U} \coloneq \{(q_j, \boldsymbol{d}_j, \boldsymbol{y}_j)\}_{j=1}^N$.

    \item \textbf{Numerical Representation Function $\eta$}

    The function $\eta$ takes as 
    input a triplet from the dataset 
    $\mathcal{U}$ and outputs the 
    corresponding numerical representation 
    that will be used to train, validate, 
    or test $f$ in practice.
    The nature of the $\eta$'s 
    output is determined by the strategy 
    employed to solve the ranking problem,
    and consequently on the learning 
    algorithm, the input and output spaces,
    and additional relevant factors.

    To illustrate the function more 
    clearly, let us consider some 
    examples. $\eta$ might take 
    as input an image query, a 
    collection of text documents, and
    a binary ground-truth with `yes' and 
    `no' labels; it could then generate 
    a vector representation of the image 
    using ResNet~\cite{He_2015_ResNet_paper}, 
    a vector representation of 
    the documents using BERT~\cite{Devlin_2019_BERT_paper}, 
    and a binary vector representing 
    relevance.
    Alternatively, $\eta$ could 
    receive a textual query, a 
    collection of textual documents, 
    and a relevance score for each 
    query-document pair; producing a
    feature vector for each pair and 
    a corresponding vector where the 
    relevance score is normalized 
    between $0$ and $1$.

    \item \textbf{Evaluation Metric $m$}

    The evaluation metric $m$ is 
    indispensable for assessing the 
    quality of rankings produced by the 
    ranking function, providing an 
    objective measure of the model's 
    performance and effectiveness on 
    the specific ranking task.
    In other words, it offers a 
    quantitative evaluation
    that enables us to understand how 
    well the model is performing.

    Various metrics exist, each 
    measuring different aspects of the 
    ranking. The choice of the most 
    appropriate metric depends on the 
    specific ranking problem.
    In the following, we provide a 
    selection of ranking metrics:
    \textit{Mean Reciprocal Rank}~\cite{Yu_2024_Evaluation_Metrics_paper},
    \textit{Mean Average Precision}~\cite{Yu_2024_Evaluation_Metrics_paper},
    \textit{Normalized Discounted Cumulative Gain}~\cite{Jarvelin_2002_NDCG_paper},
    \textit{Rank-Biased Precision}~\cite{Moffat_2008_RBP_paper}, and
    \textit{Expected Reciprocal Rank}~\cite{Chapelle_2009_ERR_paper}.

    \item \textbf{Loss Function $l$}
    
    The loss function $l$, 
    also known as the error function, 
    cost function, or objective function,
    quantifies the error between the 
    predicted value (output of $f$)
    and the true value (ground-truth).
    The loss function is instrumental in 
    training the function $f$,
    serving as a guiding objective for 
    optimization algorithms~\cite{Sun_2019_Optimization_Survey_paper}.
    During training, optimization 
    algorithms aim to minimize the 
    loss function by updating the 
    parameters of $f$,
    thereby improving the accuracy 
    of the resulting rankings.
    The mathematical formalization of 
    the loss function is as follows:
    $l(f(q, \boldsymbol{d}), \boldsymbol{y}) = l(\hat{\boldsymbol{y}}, \boldsymbol{y})$.

    In the context of ranking, minimizing 
    a loss function is equivalent to 
    maximizing a ranking metric. 
    However, the non-continuous and 
    non-differentiable nature of ranking 
    metrics presents an obstacle for 
    optimization algorithms.
    To circumvent this issue, 
    \textit{surrogate objective functions}~\cite{Bruch_2019_SurrogateObjective_paper,
    Burges_2005_SurrogateObjective_paper, 
    Cao_2007_SurrogateObjective_paper,
    Bruch_2021_CrossEntropy_paper,
    Qin_2010_SurrogateObjective_paper,
    Taylor_2008_SurrogateObjective_paper}
    are utilized. These are continuous and 
    differentiable functions derived from 
    ranking metrics. Their use allows 
    optimization algorithms to run smoothly 
    and optimize objective functions 
    consistent with ranking metrics.
 
    \item \textbf{Hypothesis Class of $f$}
    
    The hypothesis class represents the 
    family of functions to which 
    $f$ belongs.
    Hypothesis classes can be linear 
    functions, polynomial functions, 
    and, among the most commonly used 
    in LTR, decision forests and deep 
    neural networks~\cite{Bruch_2023_FTIR_book}.
    The goal is to identify the most 
    suitable hypothesis class for a 
    given problem, as using all the 
    other aforementioned ingredients 
    ($\mathcal{U}$, $\eta$, and 
    $l$ with an optimization algorithm), 
    we will obtain the optimal function 
    within that class, which will 
    be our learnt ranking function 
    $f(\cdot, \cdot; \theta)$, where 
    $\theta$ represents the learnt 
    parameters that minimize the loss 
    function.

    To make this more intuitive, 
    consider the analogy of a nail 
    in a wall. Our goal is to remove 
    this nail using a specific tool 
    ($f$). To achieve 
    this, we can employ various tools 
    (which represent the hypothesis 
    classes), such as the `class of 
    pliers', the `class of hammers', 
    the `class of screwdrivers', and 
    many others. After careful analysis, 
    we decide, based on our 
    considerations, to select the `class 
    of pliers' (the chosen hypothesis 
    class), which contains a variety of 
    pliers (family of functions), 
    including the particular pair of 
    pliers ($f$) recommended 
    by a domain expert who selected it 
    according to specific analyses.
     
\end{itemize}

\noindent Through the careful 
selection and combination of these
ingredients, it is possible to 
develop highly effective learnt 
ranking functions capable of solving 
ranking problems satisfactorily.

To gain a more comprehensive 
understanding of the learning process 
for the ranking function $f$, we revisit 
the necessary components and illustrate 
how these elements interact.

Central to our learning and evaluation 
process for function $f$ is the dataset 
$\mathcal{U} \coloneq \{(q_j, \boldsymbol{d}_j, \boldsymbol{y}_j)\}_{j=1}^N$,
which is typically partitioned into 
three subsets: training, validation, 
and test sets~\cite{Bishop_2006_ML_book}.
The training set comprises the triples  
used to train the ranking function, 
the validation set is used to evaluate 
the model's performance during training 
and to tune hyperparameters, and the 
test set is used to assess the 
performance of the final learnt ranking 
function $f$.

Once the dataset is acquired, the 
subsequent step involves applying 
a numerical representation function, 
$\eta$. This function transforms each 
dataset instance into a numerical 
representation tailored to the specific
ranking problem and chosen solution 
strategy.
The resulting representations are then 
employed to train and evaluate the 
ranking function $f$. The quality and 
characteristics of $f$ are heavily 
influenced by $\eta$.

While $\eta$ plays a vital role, the 
selection of a suitable ranking metric 
$m$ is equally important.
A ranking metric is essential for 
objectively evaluating the performance 
of a ranking function.
Our goal is to learn a function $f$ 
that maximizes the chosen ranking 
metric, which is achieved by minimizing
a surrogate objective function $l$ that
is correlated with the metric to 
be maximized. The function $l$ is then
employed to train $f$.

The final step prior to training the
ranking function $f$ involves selecting 
the hypothesis class. This choice 
defines the space of possible functions 
from which the learnt ranking function 
will be extracted.
It is important to note that, similar 
to the selection of the numerical 
representation function $\eta$, 
the ranking metric $m$, and the loss 
function $l$, there is no universally 
optimal choice for the hypothesis 
class. Instead, this selection requires 
careful consideration of the specific 
ranking problem.

Having prepared all the necessary 
components, we can now proceed with 
the actual training of the ranking 
function $f$.
Specifically, we utilize the training 
set and validation set, appropriately 
transformed by $\eta$, along with the 
loss function $l$ and an optimization
algorithm~\cite{Sun_2019_Optimization_Survey_paper}.
The training process involves 
optimizing the parameters of 
$f(\cdot, \cdot; \theta)$ using 
an optimization algorithm to 
minimize the loss function on the 
training set.
Beyond the training set, which 
is essential for parameter estimation,
the validation set plays a pivotal 
role for hypothesis (function) 
selection.
By assessing the predictive performance
of different functions within the 
hypothesis class on the validation set, 
we can determine the function that is 
most likely to generalize well to new, 
unseen data, thereby enabling us to 
select the optimal learnt ranking 
function $f$.

Once the final learnt ranking function, 
$f$, has been obtained, it is evaluated 
on the test set using the selected metric,
$m$, to determine its effectiveness in 
solving the given ranking task.

\chapter{Learning Cluster Representatives}

In this chapter, we delve into the 
details of our novel methodology 
for MIPS-based clustering-based ANN 
search, which represents the 
state-of-the-art in the field.
This methodology was presented 
at SIGIR 2024\footnote{\url{https://sigir-2024.github.io/}}
in paper~\cite{Vecchiato_2024_paper}.
In this thesis, we expand upon 
\cite{Vecchiato_2024_paper} and provide 
a more in-depth analysis. We begin by 
exploring the intuition and motivations 
behind our research. Subsequently, we 
formalize the methodology rigorously, 
outlining its fundamental principles and 
components. We then present the 
experimental setup and discuss the 
obtained results, extending beyond those 
reported in the aforementioned paper.
Finally, we dedicate a section to further 
analysis, where we explore a variation 
of the methodology and evaluate its 
corresponding results.

\section{Intuition}

In Section 3.1 on clustering-based ANN 
search, we saw that at query time, 
the fundamental element analyzed by the 
routing function $\tau$, to return the 
$\ell$ most similar partitions to a 
given query, is the representative element 
of each cluster. The standard approach 
consists of computing the mean of the 
points within the cluster as the 
representative point (Equation~\ref{equation:tau-mips}).

However, a natural question arises:
Does there exist a representative point 
that better captures the semantics of 
the cluster and its internal elements, 
such that when a distance function $\delta$ 
is applied, the returned score more 
accurately reflects the true similarity 
between the query and the cluster?
Our work provides an affirmative answer 
to this question by introducing 
\emph{learnt representative points}.

\begin{figure}
    \centering
    \includegraphics[width=1\linewidth]{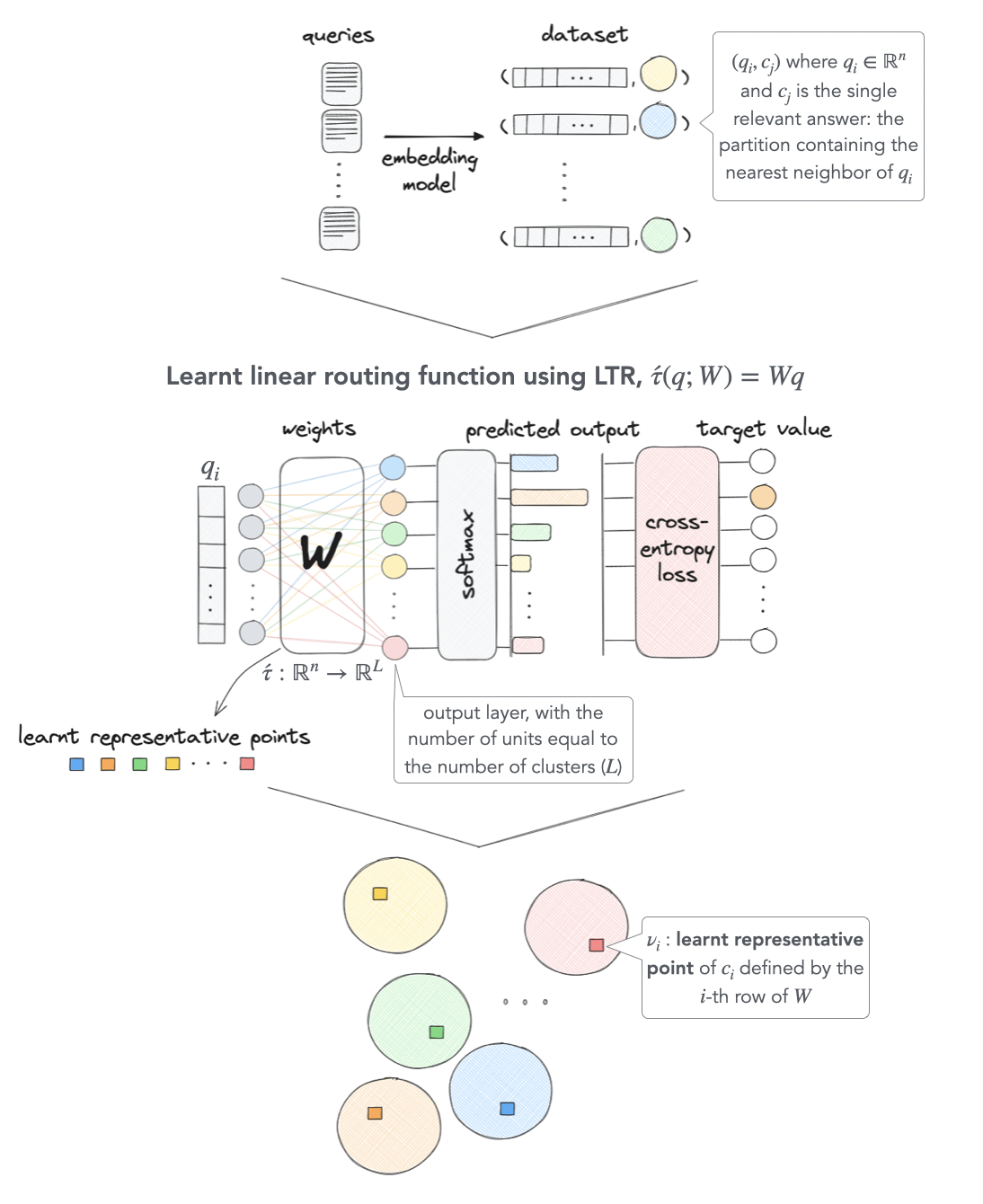}
    \caption{\footnotesize
    Visual representation of the 
    learning process of the routing 
    function $\acute{\tau}$ using LTR 
    to obtain the learnt representative 
    points. Specifically, three phases 
    are depicted: dataset creation, 
    learning the routing function,
    and visualization of clusters with 
    the learnt points.
    }
    \label{fig:our-method}
\end{figure}

Let us begin with a simple yet crucial 
observation to understand how the 
learnt representative points are obtained:
The routing function $\tau$ is a ranking 
function that addresses the problem of 
ordering partitions from the most similar, 
which has the highest probability of 
containing the actual top-$k$ documents 
for a given query, to the least similar,
where we have the lowest probability of 
finding the resulting top-$k$ documents.

This observation implies that the 
function $\tau$ can be learnt using 
Learning-to-Rank (Section 4), and in 
particular, we can learn it using a 
simple linear function $\acute{\tau}(q;\; W) = Wq$,  
with $q \in \mathbb{R}^n$ and 
$W \in \mathbb{R}^{L \times n}$,
where the $i$-th row of $W$ is the learnt 
representative vector of the $i$-th
partition. This yields a learnt ranking 
function $\acute{\tau}(q;\; W) = Wq$ 
that, given a query $q$, outputs a ranking 
score for each of the $L$ partitions 
under consideration, with $W$ 
containing the new representatives of 
the partitions.

Concretely, to learn the routing function 
$\acute{\tau}$, we consider the top-$k$ 
case with $k = 1$. 

It is straightforward 
to observe that the quality of routing can 
be evaluated using a ranking metric: Given 
a query, the routing function assigns to 
each partition a likelihood of containing 
the nearest neighbor, while the single 
relevant response (ground-truth) is the 
partition that actually contains the nearest
neighbor. An appropriate metric we can use 
is MRR~\cite{Yu_2024_Evaluation_Metrics_paper}.
Consequently, with MRR as the appropriate 
metric and since each query has a single 
relevant answer, the most suitable surrogate 
objective to optimize is cross-entropy 
loss~\cite{Bruch_2021_CrossEntropy_paper, 
Bruch_2019_CrossEntropy_paper}.

Regarding the dataset, it is also easily 
obtainable. All that is needed is a set 
of queries where, for each query, as 
ground-truth relevance label, we have the 
corresponding partition that contains its 
nearest neighbor. The corresponding 
partition can be found by simply retrieving 
the relevant top-$1$ document with an 
exhaustive search and subsequently identifying
its cluster of belonging.

Finally, by representing the query as 
a vector in $\mathbb{R}^n$ and the 
ground-truth as a binary vector with 
a $1$ set for the partition containing 
the top-$1$ document, and selecting a 
hypothesis class of linear neural 
networks~\cite{Goodfellow_2016_DL_book}
with only an input layer (with the number 
of units equal to the dimension of the
query vector) and an output layer 
(with units equal to the number of 
clusters) followed by with a softmax
function, we have all the necessary
components to learn $\acute{\tau}$, 
where the learnt representative points 
in this case are the weights $W$ of 
the neural network.
A graphical illustration of the 
aforementioned is presented in 
Figure~\ref{fig:our-method}.

Building upon this intuition, we 
have experimentally demonstrated that 
the learnt representative points 
significantly improve the 
accuracy of clustering-based ANN search.
Moreover, an appealing aspect of 
this methodology lies in its easy 
integration into a production system 
that already implements a routing 
mechanism. By merely replacing the 
old representative points with the 
learnt points, without any additional
modifications.
\section{Methodology}

In this section, we formally introduce 
the methodology employed to learn the 
$L$ representative points, where $L$ 
denotes the number of clusters. We provide
a rigorous mathematical formulation, 
detailing the steps involved, starting 
from the problem to be solved.

Given a collection of data points 
$\mathcal{X} \subset \mathbb{R}^n$
and a query $q \in \mathbb{R}^n$, 
obtained from an embedding function 
$\varphi$, we aim to retrieve the 
top-$k$ most similar data points
to $q$, as measured by the inner product,
thereby solving the $k$-MIPS problem
(Definition~\ref{definition:k-MIPS}):
$\{u^\ast_1, u^\ast_2, \cdots,  u^\ast_k\} \subseteq \argmax_{u \in \mathcal{X}}^{(k)} \langle q, u \rangle$ 
(Equation~\ref{equation:k-MIPS}).
As shown in Section 3.1, an efficient 
practical solution for this problem, though 
approximate, involves employing 
clustering-based ANN search approach.

Clustering-based ANN search applies 
a clustering algorithm 
$\mathcal{C}: \mathbb{R}^n \rightarrow \{1, 2, \dots, L\}$
to partition the dataset $\mathcal{X}$ 
into a set of $L$ non-overlapping 
clusters $\{c_i\}_{i=1}^L$, where 
each cluster is represented by a
representative point $\mu_i$. 
In the standard approach, $\mu_i$ 
is computed as the mean of the points 
in cluster $c_i$. Subsequently, given 
a query $q$, a routing function 
$\tau: \mathbb{R}^n \rightarrow \argmax^{\ell}\mathbb{R}^L$
retrieves the top-$\ell$ clusters.
To retrieve the top-$\ell$ clusters, 
$\tau$ first assigns a similarity score
to each cluster by computing the inner 
product between $q$ and each $\mu_i$, and 
then returns the $\ell$ clusters with 
the highest scores. 

Mathematically, $\tau$ can be represented as 
$\tau(q;\; M) =  \argmax_{i = 1, \dots, L}^{(\ell)} (Mq), $
where $q \in \mathbb{R}^n$ and 
$M \in \mathbb{R}^{L \times n}$ whose 
$i$-th row is $\mu_i$.
The search space for the final 
top-$k$ documents will be 
restricted to points contained 
in the top-$\ell$ clusters.

In this work, we propose to replace 
$\{\mu_i\}_{i=1}^L$ with 
learnt representative points 
$\{\nu_i\}_{i=1}^L$.
To this end, we learn a routing 
function $\acute{\tau}(q;\; W) = Wq$,
where $W \in \mathbb{R}^{L \times n}$, 
using LTR. $W$ will substitute $M$ 
in the original formulation.

In order to learn the linear function 
$W$, the training dataset consists of 
a set of pairs: $(q, (b_i)_{i=1}^L)$.
Here, $q$ is a query point 
$\in \mathbb{R}^n$, and $(b_i)_{i=1}^L$
is a binary vector of length equal to 
the number of clusters, with exactly 
one element set to $1$.
The $1$ is located in the $i$-th 
position corresponding to the $i$-th 
cluster, $c_i$, that contains the 
top-$1$ document for query $q$.
$(b_i)_{i=1}^L$ represents the 
ground-truth for $q$.

To train $\acute{\tau}$ another 
necessary ingredient is to define a 
loss function. Given that we aim to 
maximize the Mean Reciprocal Rank (MRR),
a metric well-suited for this task, 
the cross-entropy loss emerges as an 
appropriate choice. Notably, prior research, 
as demonstrated in \cite{Bruch_2021_CrossEntropy_paper} 
and \cite{Bruch_2019_CrossEntropy_paper}, 
has established cross-entropy loss as 
a consistent surrogate for MRR under the 
assumption that each query has at most 
one relevant item with probability $1$,  
as is the case in our setting.

For completeness, we provide the 
mathematical definitions of MRR and 
cross-entropy loss for a single 
query $q$. 
MRR is given by:
\begin{equation}
    \label{equation:MRR}
    \text{MRR}(\boldsymbol{q}) = \frac{1}{|\boldsymbol{q}|} \sum_{i=1}^{|\boldsymbol{q}|} \frac{1}{r(\boldsymbol{q}_i)},
\end{equation}
in which $|\boldsymbol{q}|$ is the 
cardinality of the query set, and 
$r(\boldsymbol{q}_i)$ represents the 
rank position of the first relevant item 
for the query $\boldsymbol{q}_i$.
The cross-entropy loss (CE) for a
single query $q$ is computed as:
\begin{equation}
    \label{equation:l-CE}
    l_\text{CE}(\boldsymbol{s}, \boldsymbol{b}) = 
    - \sum_{i=1}^L \boldsymbol{b}_i \log \text{softmax}(\boldsymbol{s})_i = 
    - \sum_{i=1}^L \boldsymbol{b}_i \log \frac{\exp({\boldsymbol{s}_i)}}{\sum_{j=1}^L \exp{(\boldsymbol{s}_j)}},
\end{equation}
whereby $\boldsymbol{s} = Wq = \acute{\tau}(q;\; W)$
and $\boldsymbol{b} = (b_i)_{i=1}^L$.

Specifically, our final objective 
function, which we aim to minimize 
to learn the function $\acute{\tau}$,
is the cross-entropy loss across the 
entire training query set, defined as:
\begin{equation}
    \label{equation:L-CE}
    \mathcal{L}_\text{CE}(S, B) = \frac{1}{N} \sum_{i=1}^N l_\text{CE}(S_i, B_i),
\end{equation}
where $S = (\boldsymbol{s}_i)_{i=1}^N = (Wq_i)_{i=1}^N$ 
and $B = (\boldsymbol{b}_i)_{i=1}^N$.
$N$ is the number of samples in 
the training query set.

With the training query set and 
cross-entropy loss in place, coupled with 
the selection of an optimizer, we are now 
equipped to learn $\acute{\tau}$ and 
obtain the learnt representative points 
$\{\nu_i\}_{i=1}^L$.

\subsection{Generalizing to Top-$k$}

The foregoing provides a formal description
of how to learn a function that, given a 
query, accurately identifies the cluster 
containing the top-$1$ document. This 
approach will be used throughout our 
experiments, demonstrating satisfactory 
performance not only in the $1$-MIPS case 
but also in the $10$-MIPS and $100$-MIPS 
scenarios.

However, generalizing the learning approach
for the routing function $\acute{\tau}$ to the 
top-$k$ problem with $k > 1$ is straightforward.
We will now demonstrate why, by defining the 
training query set and loss function for this 
context.

The training query set consists of a 
collection of pairs: $(q, (b_i)_{i=1}^L)$,  
where $q$ represents a query, and $(b_i)_{i=1}^L$
is a binary vector. The $i$-th position of 
$(b_i)_{i=1}^L$ is set to $1$ if and only 
if the $i$-th cluster contains at least one
of the top-$k$ documents relevant to $q$.
Formally, given a set $\mathcal{S}$ 
containing the top-$k$ data points for 
query $q$, the set of clusters that 
contain at least one element of 
$\mathcal{S}$ is defined as:
$\{c_i \, | \, c_i \in \{c_i\}_{i=1}^L, \, c_i \cap \mathcal{S} \neq \emptyset\}$.

Regarding the cross-entropy loss function 
over the entire training query set, it 
can be updated following the formulation 
outlined in~\cite{Bruch_2021_CrossEntropy_paper} as follows:
 
\begin{equation}
    \label{equation:L-CE-k}
    \mathcal{L}_{\text{CE}}^{\text{top}-k}(S, B) = - \frac{1}{N} \sum_{i=1}^N \sum_{j=1}^L \frac{2^{B_{ij}} - \gamma_{ij}}{\sum_{z=1}^L 2^{B_{iz}} - \gamma_{iz}} \log \frac{\exp(S_{ij})}{\sum_{z=1}^L \exp(S_{iz})}
\end{equation}
with $S$, $B$ and $N$ as defined in  
$\mathcal{L}_\text{CE}(S, B)$, and 
$\gamma$'s sampled uniformly from $\left[0, 1\right]$.

With this formulation, it is possible 
to learn a function $\acute{\tau}$ that,
given a query, is trained to identify the 
partitions containing the top-$k$ documents, 
thus generalizing from the top-$1$ training 
case.

\section{Experimental Setup}

The following section presents the 
experimental setup used in this work.

\subsubsection{Datasets}
Three publicly available datasets were 
utilized for this research\footnote{\url{https://ir-datasets.com}}.
\begin{itemize}
    \item \textsc{MS MARCO}~\cite{Nguyen_2016_MsMarco_paper}:
    Large scale MAchine Reading 
    COmprehensive dataset focused 
    on deep learning in search.
    It comprises of $8{,}841{,}823$
    short passages and $909{,}824$ 
    train queries.

    \item \textsc{HotpotQA}~\cite{Yang_2018_HotpotQA_paper}:
    Large scale Wikipedia-based question
    answering dataset, comprising of 
    $5{,}233{,}235$ documents and 
    $97{,}852$ queries.

    \item \textsc{FEVER}~\cite{Thorne_2018_Fever_paper}:
    Fact Extraction and VERification
    dataset consists of $5{,}396{,}138$ 
    documents and $13{,}332$ queries.
\end{itemize}

\subsubsection{Embedding Models}
To transform queries and documents into 
$n$-dimensional dense vectors, we employed 
embedding models. Specifically, we utilized 
the following Transformer-based 
Pre-trained Language Models (PLMs).

\begin{itemize}
    \item \texttt{TAS-B}\footnote{\url{https://huggingface.co/sentence-transformers/msmarco-distilbert-base-tas-b}}~\cite{Hofstatter_2021_TASB_paper}:
    The model projects text inputs into 
    $768$-dimensional dense vectors. 
    A distinguishing feature of this model 
    is the employment of balanced topic 
    aware sampling.
    
    \item \texttt{Contriever}\footnote{\url{https://huggingface.co/facebook/contriever}}~\cite{Izacard_2022_Contriever_paper}:
    The model outputs $768$-dimensional 
    dense vector representations of the 
    input texts. Contrastive learning is 
    the principal distinguishing technique
    in the model.

    \item \texttt{all-MiniLM-L6-v2}\footnote{\url{https://huggingface.co/sentence-transformers/all-MiniLM-L6-v2}}, \texttt{all-mpnet-base-v2}\footnote{\url{https://huggingface.co/sentence-transformers/all-mpnet-base-v2}} and \texttt{all-distilroberta-v1}\footnote{\url{https://huggingface.co/sentence-transformers/all-distilroberta-v1}}~\cite{Reimers_2019_SBERT_paper}:
    Sentence transformer models that map
    text into dense vector spaces of 
    dimensionality $384$, $768$, and $768$,
    respectively. These models were 
    obtained by fine-tuning a pre-trained 
    base model\footnote{\url{https://www.sbert.net/docs/sentence_transformer/pretrained_models.html}}
    on a massive and diverse dataset 
    comprising over $1$ billion text 
    pairs.
\end{itemize}
 
\subsubsection{Baseline}
The baseline used for comparison with 
our learnt routing function is defined 
as follows: $\tau(q) = Mq$, where the 
rows of $M$ represent the centroids 
of the clusters returned by the 
clustering algorithm.
In other words, to assess the 
effectiveness of our strategy, 
we compare the learnt representative 
points $\{\nu_i\}_{i=1}^L$ with 
the standard representative points
$\{\mu_i\}_{i=1}^L$.

\subsubsection{Implementation Details}
The routing function $\acute{\tau}$ is 
learnt via a linear neural network with 
a single input and output layer, 
no biases, and a final softmax activation 
function. The input layer has a 
dimensionality equal to that of the 
vector space in which the query and document 
points reside, while the output layer has a 
dimensionality equal to the number of 
clusters. The softmax activation function 
transforms the network's output into 
probability scores, allowing for the 
computation of the cross-entropy loss 
$l_\text{CE}$, Equation~\ref{equation:l-CE}.
To minimize the loss function, the 
Adam~\cite{Kingma_2017_Adam_paper} 
optimizer is employed with a
learning rate of $10^{-4}$.
A batch size of $512$ is used, and 
training is conducted for a maximum 
of $100$ epochs.
Upon training completion, the network 
weights $W$ correspond to the learnt 
representative points, resulting in the 
learnt routing function 
$\acute{\tau}(q; W) = Wq$,
with $W \in \mathbb{R}^{L \times n}$.

The pipeline for obtaining learnt 
representative points and evaluating 
their effectiveness is as follows.
First, a dataset and an embedding model 
are employed to transform queries and 
documents within the dataset into 
vector representations using the given 
embedding. Subsequently, a clustering 
algorithm is applied to partition the 
set of document vector points 
$\{d_i\}_{i=1}^P \subset \mathbb{R}^n$ 
into $L = \sqrt{P}$ clusters. In this
study, we consider Standard KMeans, 
Spherical KMeans (implemented using 
the FAISS library~\cite{Douze_2024_FAISS_paper}), 
and Shallow KMeans. Following clustering, 
for each query $q_i \in \mathbb{R}^n$ in
the query set $\{q_i\}_{i=1}^N \subset \mathbb{R}^n$,
we obtain the pair $(q_i, \boldsymbol{b}_i)$ 
as described in Section 5.2. The resulting
set of pairs is divided into training
($60\%$), validation ($20\%$), and test 
($20\%$) sets. Finally, our neural network
is trained on the training set, with 
the best model selected based on the loss
function on the validation set.
The learnt representative vectors 
$\{\nu_i\}_{i=1}^L$ and the standard 
representative points $\{\mu_i\}_{i=1}^L$
are evaluated on the test set.

\subsection{Evaluation Metric}
The learnt representative points 
$\{\nu_i\}_{i=1}^L$ and 
standard representative points 
$\{\mu_i\}_{i=1}^L$ are evaluated 
in terms of \emph{top-$k$ accuracy}.
Top-$k$ accuracy is obtained by  
retrieving the top-$\ell$ partitions 
for each query $q$ based on the 
routing function, i.e., $\tau(q) = Mq$ 
for the baseline and $\acute{\tau}(q) = Wq$ 
for our method, and then computing the 
percentage of the top-$k$ documents 
contained within these top-$\ell$ 
partitions.

\section{Experimental Results}
In this section, we present the 
experimental results obtained by 
evaluating the performance of our 
proposed method, \emph{Learnt}, against 
the standard clustering-based ANN 
search, \emph{Baseline}.
The evaluation is conducted on the 
datasets, embedding models, and 
clustering algorithms discussed in the 
previous section using the top-$k$ 
accuracy metric.

\subsection{Top-$1$ Retrieval}
\setlength{\tabcolsep}{8pt}
\renewcommand{\arraystretch}{1.2}
\begin{table}[b]
    \centering
    \caption{\footnotesize
    Top-$1$ accuracy of Baseline and Learnt
    methods using \texttt{all-MiniLM-L6-v2} 
    on \textsc{MS MARCO}, \textsc{HotpotQA},
    and \textsc{FEVER} datasets.
    Results are reported for Standard, Spherical,
    and Shallow KMeans clustering algorithms, 
    considering top-$\ell$ partitions with 
    $\ell$ equal to $0.1\%$ and $1\%$ of $L$.
    }
    \footnotesize
    \footnotesize
    \begin{tabular}{l l cc cc cc}
        \toprule
        \multirow{2}{*}{Dataset} & \multirow{2}{*}{Method} & \multicolumn{2}{c}{Standard} & \multicolumn{2}{c}{Spherical} & \multicolumn{2}{c}{Shallow} \\
        
        & & 0.1\% & 1\% & 0.1\% & 1\% & 0.1\% & 1\% \\
        \toprule
        
        \multirow{2}{*}{\textsc{MS MARCO}}
        & {Baseline} & $0.392$ & $0.779$ & $0.627$ & $0.869$ & $0.517$ & $0.815$\\
        & {Learnt}   & $0.746$ & $0.940$ & $0.751$ & $0.938$ & $0.670$ & $0.923$\\
        \midrule
        
        \multirow{2}{*}{\textsc{HotpotQA}}
        & {Baseline} & $0.089$ & $0.481$ & $0.328$ & $0.684$ & $0.258$ & $0.724$\\
        & {Learnt}   & $0.488$ & $0.844$ & $0.493$ & $0.833$ & $0.412$ & $0.827$\\
        \midrule

        \multirow{2}{*}{\textsc{FEVER}}
        & {Baseline} & $0.102$ & $0.443$ & $0.249$ & $0.562$ & $0.279$ & $0.621$\\
        & {Learnt}   & $0.663$ & $0.865$ & $0.662$ & $0.872$ & $0.633$ & $0.912$\\
        \bottomrule
    \end{tabular}
    \label{tab:allmini-top-1}
\end{table}
\begin{table}
    \centering
    \caption{\footnotesize
    Top-$1$ accuracy of Baseline and Learnt
    methods using \texttt{TAS-B} 
    on \textsc{MS MARCO}, \textsc{HotpotQA},
    and \textsc{FEVER} datasets.
    Results are reported for Standard, Spherical, 
    and Shallow KMeans clustering algorithms, 
    considering top-$\ell$ partitions with 
    $\ell$ equal to $0.1\%$ and $1\%$ of $L$.
    }
    \footnotesize
    \begin{tabular}{l l cc cc cc}
        \toprule
        \multirow{2}{*}{Dataset} & \multirow{2}{*}{Method} & \multicolumn{2}{c}{Standard} & \multicolumn{2}{c}{Spherical} & \multicolumn{2}{c}{Shallow} \\
        
        & & 0.1\% & 1\% & 0.1\% & 1\% & 0.1\% & 1\% \\
        \toprule
        
        \multirow{2}{*}{\textsc{MS MARCO}}
        & {Baseline} & $0.480$ & $0.835$ & $0.680$ & $0.915$ & $0.553$ & $0.869$\\
        & {Learnt}   & $0.727$ & $0.936$ & $0.724$ & $0.933$ & $0.612$ & $0.896$\\
        \midrule
        
        \multirow{2}{*}{\textsc{HotpotQA}}
        & {Baseline} & $0.258$ & $0.724$ & $0.372$ & $0.783$ & $0.345$ & $0.756$\\
        & {Learnt}   & $0.525$ & $0.882$ & $0.507$ & $0.867$ & $0.405$ & $0.797$\\
        \midrule

        \multirow{2}{*}{\textsc{FEVER}}
        & {Baseline} & $0.235$ & $0.675$ & $0.314$ & $0.690$ & $0.303$ & $0.704$\\
        & {Learnt}   & $0.836$ & $0.930$ & $0.834$ & $0.925$ & $0.819$ & $0.917$\\
        \bottomrule
    \end{tabular}
    \label{tab:tasb-top-1}
\end{table}

Tables~\ref{tab:allmini-top-1} and 
\ref{tab:tasb-top-1} represent the 
top-$1$ accuracy of the Baseline 
method, with standard representative 
points $\{\mu_i\}_{i=1}^L$, and the 
Learnt method, with learnt 
representative points 
$\{\nu_i\}_{i=1}^L$, on the 
\textsc{MS MARCO}, \textsc{HotpotQA}, 
and \textsc{FEVER} text datasets.
Table~\ref{tab:allmini-top-1} employs 
the \texttt{all-MiniLM-L6-v2}, while 
Table~\ref{tab:tasb-top-1} uses the 
\texttt{TAS-B} embedding. 
Standard, Spherical, and Shallow KMeans 
clustering algorithms are employed to 
generate partitions. For each algorithm, 
top-$1$ accuracy is measured considering 
the top $0.1\%$ and $1\%$ of clusters 
relative to the total number, i.e., 
$\ell = 0.1\% \times L$ and 
$\ell = 1\% \times L$.

Tables~\ref{tab:allmini-top-1} 
and \ref{tab:tasb-top-1} 
clearly demonstrate that our 
proposed Learnt method consistently 
outperforms the Baseline. 
For instance, considering Standard 
KMeans with $\ell = 0.1\% \times L$,
$\{\nu_i\}_{i=1}^L$ exhibit a 
significant accuracy improvement 
compared to $\{\mu_i\}_{i=1}^L$, 
achieving $+90.3\%$ on \textsc{MS MARCO}, 
$+448.3\%$ on \textsc{HotpotQA}, and 
$+550.0\%$ on \textsc{FEVER} when using 
\texttt{all-MiniLM-L6-v2}, and 
 $+51.5\%$ on \textsc{MS MARCO}, 
$+103.5\%$ on \textsc{HotpotQA}, and 
$+255.7\%$ on \textsc{FEVER} when 
using \texttt{TAS-B}.

Additionally, analysis of the tables 
reveals a more pronounced difference 
in accuracy for $\ell = 0.1\% \times L$ 
compared to $\ell = 1\% \times L$. 
Generally, our results indicate that a 
smaller value of $\ell$ correlates with 
a larger gap in accuracy between the 
two methods, indicating that partitions 
closer to query points are of much higher 
quality. This is intuitively expected, as 
our learning objective is specifically 
tailored to identify and retrieve the 
top-$1$ cluster.

Another notable observation from the 
results is that, across clustering 
algorithms, the accuracy difference 
between the Learnt and Baseline methods
is most marked for Standard KMeans, 
followed by Shallow KMeans, and is 
least marked for Spherical KMeans.
These findings are not unexpected for 
Standard and Spherical KMeans, considering
their distance metric used in the 
objective function. Standard KMeans utilizes 
the Euclidean distance
(Equation~\ref{equation:euclidean-distance}),
whereas Spherical 
KMeans utilizes the cosine distance 
(Equation~\ref{equation:cosine-distance}), 
which is more suitable for a MIPS-based 
problem.
On the other hand, interesting results 
are obtained for Shallow KMeans which 
performs remarkably well, despite its 
intrinsic simplicity 
(see Algorithm~\ref{alg:shallow-kmeans}).

\begin{table}
    \centering
    \caption{\footnotesize
    Top-$1$ accuracy of Baseline and Learnt
    methods on the \textsc{MS MARCO} dataset
    using \texttt{Contriever}, 
    \texttt{all-mpnet-base-v2}, 
    and \texttt{all-distilroberta-v1} 
    embedding models. 
    Results are reported for Standard, Spherical, 
    and Shallow KMeans clustering algorithms, 
    considering top-$\ell$ partitions with 
    $\ell$ equal to $0.1\%$ and $1\%$ of $L$.
    }
    \footnotesize
    \begin{tabular}{l l cc cc cc}
        \toprule
        \multirow{2}{*}{Encoding} & \multirow{2}{*}{Method} & \multicolumn{2}{c}{Standard} & \multicolumn{2}{c}{Spherical} & \multicolumn{2}{c}{Shallow} \\
        
        & & 0.1\% & 1\% & 0.1\% & 1\% & 0.1\% & 1\% \\
        \toprule
        
        \multirow{2}{*}{\texttt{Contriever}}
        & {Baseline} & $0.602$ & $0.895$ & $0.756$ & $0.938$ & $0.640$ & $0.909$\\
        & {Learnt}   & $0.790$ & $0.952$ & $0.780$ & $0.946$ & $0.690$ & $0.927$\\
        \midrule
        
        \multirow{2}{*}{\texttt{all-mpnet-base-v2}}
        & {Baseline} & $0.763$ & $0.952$ & $0.794$ & $0.958$ & $0.691$ & $0.940$\\
        & {Learnt}   & $0.818$ & $0.967$ & $0.819$ & $0.966$ & $0.733$ & $0.951$\\
        \midrule

        \multirow{2}{*}{\texttt{all-distilroberta-v1}}
        & {Baseline} & $0.752$ & $0.955$ & $0.779$ & $0.960$ & $0.664$ & $0.935$\\
        & {Learnt}   & $0.807$ & $0.967$ & $0.806$ & $0.966$ & $0.706$ & $0.945$\\
        \bottomrule
    \end{tabular}
    \label{tab:encodings-top-1}
\end{table}

To further validate the effectiveness of 
our proposed method, Table~\ref{tab:encodings-top-1}
presents additional results on top-$1$ 
accuracy using various embedding models, 
namely \texttt{Contriever}, 
\texttt{all-mpnet-base-v2}, and 
\texttt{all-distilroberta-v1}, on 
\textsc{MS MARCO}, with $\ell$ set to 
$0.1$ and $1$ percent of the total number
of clusters.

The results in Table~\ref{tab:encodings-top-1} 
reaffirm our previous findings, including the
key point that the Learnt method consistently 
outperforms the Baseline across all experimental 
settings.

When comparing Tables~\ref{tab:tasb-top-1} and 
\ref{tab:encodings-top-1}, which employ 
$768$-dimensional embeddings, the accuracy gap 
between Learnt and Baseline is less pronounced 
compared to Table~\ref{tab:allmini-top-1}, where 
$384$-dimensional embeddings are used. This suggests 
that as the dimensionality of the embedding space 
increases, the Learnt method may face challenges in 
identifying significantly better representative 
points. A more in-depth investigation of this 
phenomenon is warranted in future research.

Finally, the McNemar's 
test~\cite{McNemar_1947_test_paper} was 
conducted on all results, revealing a highly 
statistically significant difference between 
the Learnt and Baseline methods ($p$-value 
$< 0.001$), strongly supporting the superior 
performance of the learnt representative 
points.

\subsection{Top-$k$ Retrieval}

\begin{figure}[b]
    \centering
    \begin{subfigure}[b]{0.325\textwidth}
        \centering
        \includegraphics[width=\textwidth]{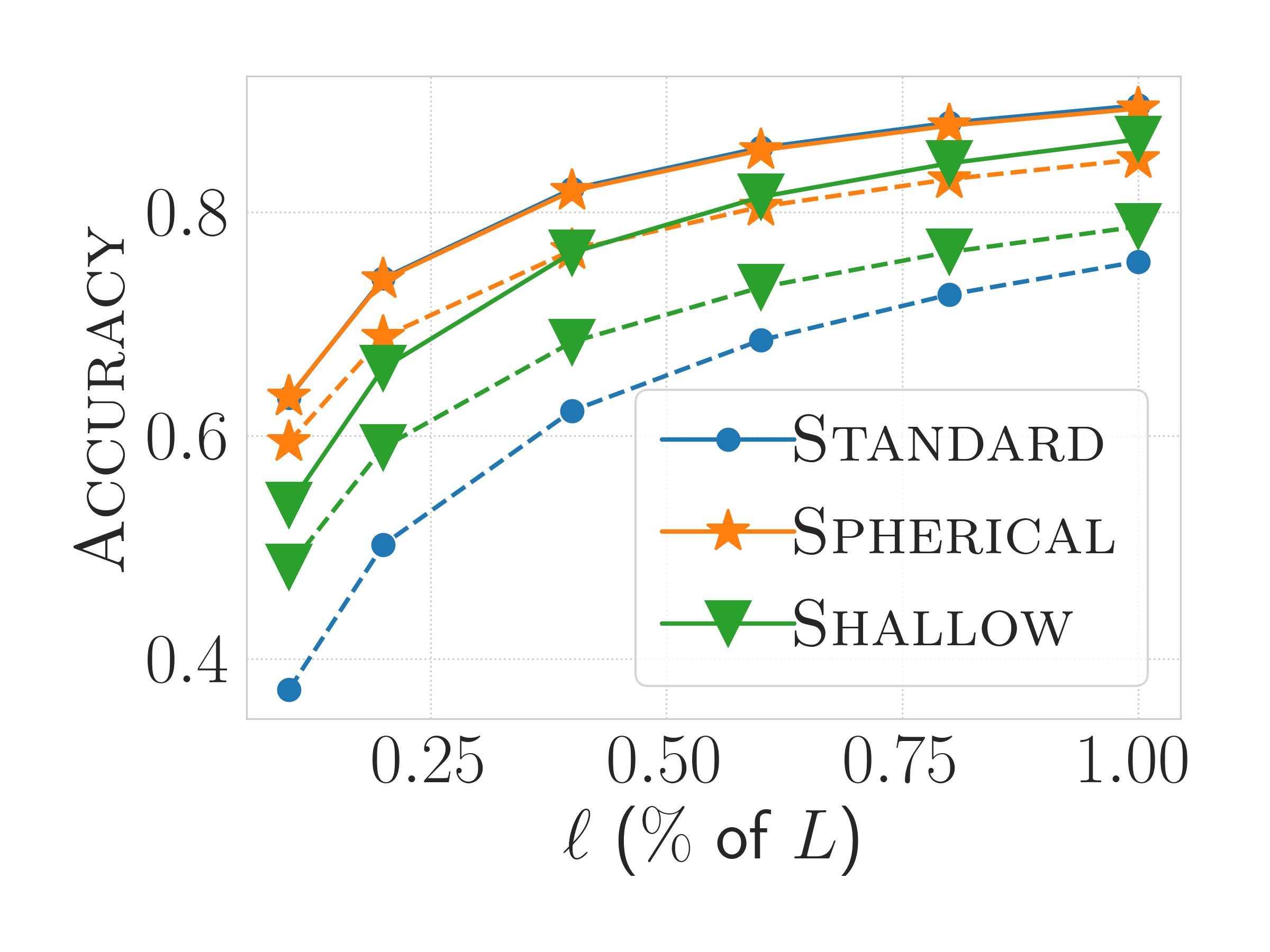}
        \caption{\footnotesize \textsc{MS MARCO}}
        \label{fig-a:top-10-msmarco}
    \end{subfigure}
    \hfill
    \begin{subfigure}[b]{0.325\textwidth}
        \centering
        \includegraphics[width=\textwidth]{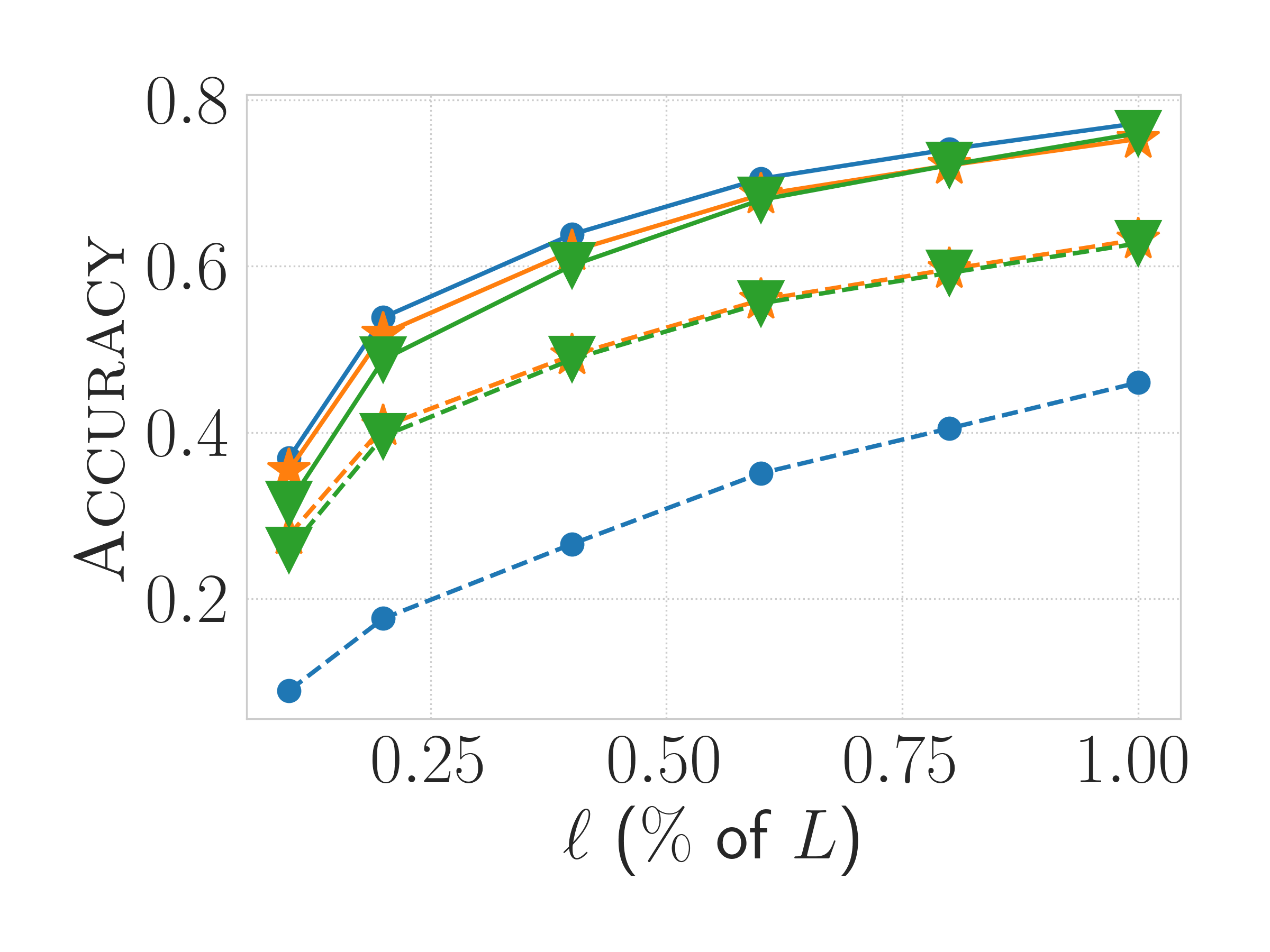}
        \caption{\footnotesize \textsc{HotpotQA}}
        \label{fig-b:top-10-hotpotqa}
    \end{subfigure}
    \hfill
    \begin{subfigure}[b]{0.325\textwidth}
        \centering
        \includegraphics[width=\textwidth]{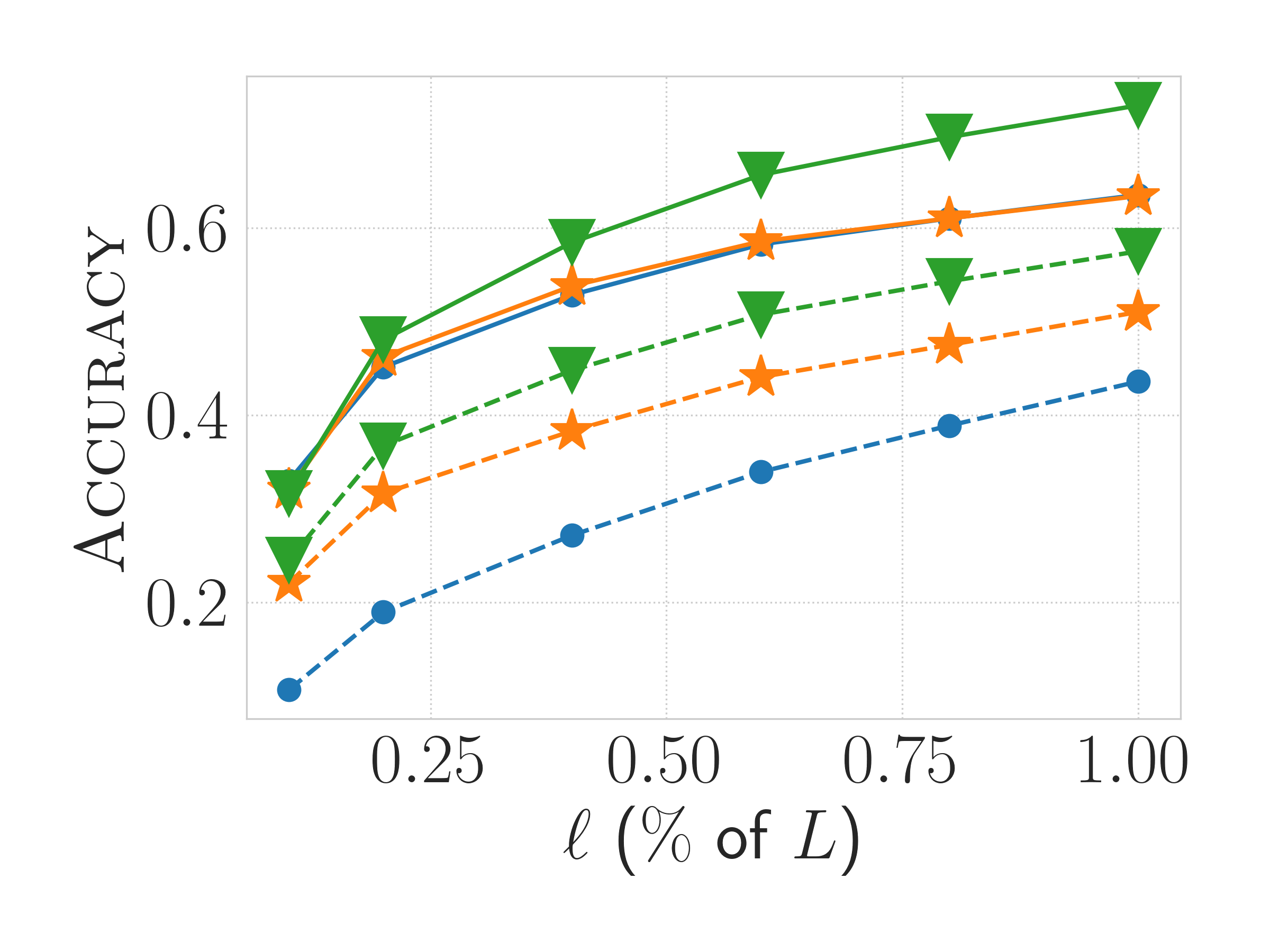}
        \caption{\footnotesize \textsc{FEVER}}
        \label{fig-c:top-10-fever}
    \end{subfigure}
    \caption{\footnotesize
    Top-$10$ accuracy of Learnt and 
    Baseline methods on \textsc{MS MARCO}, 
    \textsc{HotpotQA}, and \textsc{FEVER} 
    datasets using the \texttt{all-MiniLM-L6-v2} 
    embedding model. The $x$-axis 
    represents $\ell$, expressed as a 
    percentage of $L$,  while the 
    $y$-axis represents the accuracy.
    Solid lines indicate the Learnt 
    method, and dashed lines indicate
    the Baseline method.
    }
    \label{fig:top-10-results}
\end{figure}
\begin{figure}[t]
    \centering
    \begin{subfigure}[b]{0.325\textwidth}
        \centering
        \includegraphics[width=\textwidth]{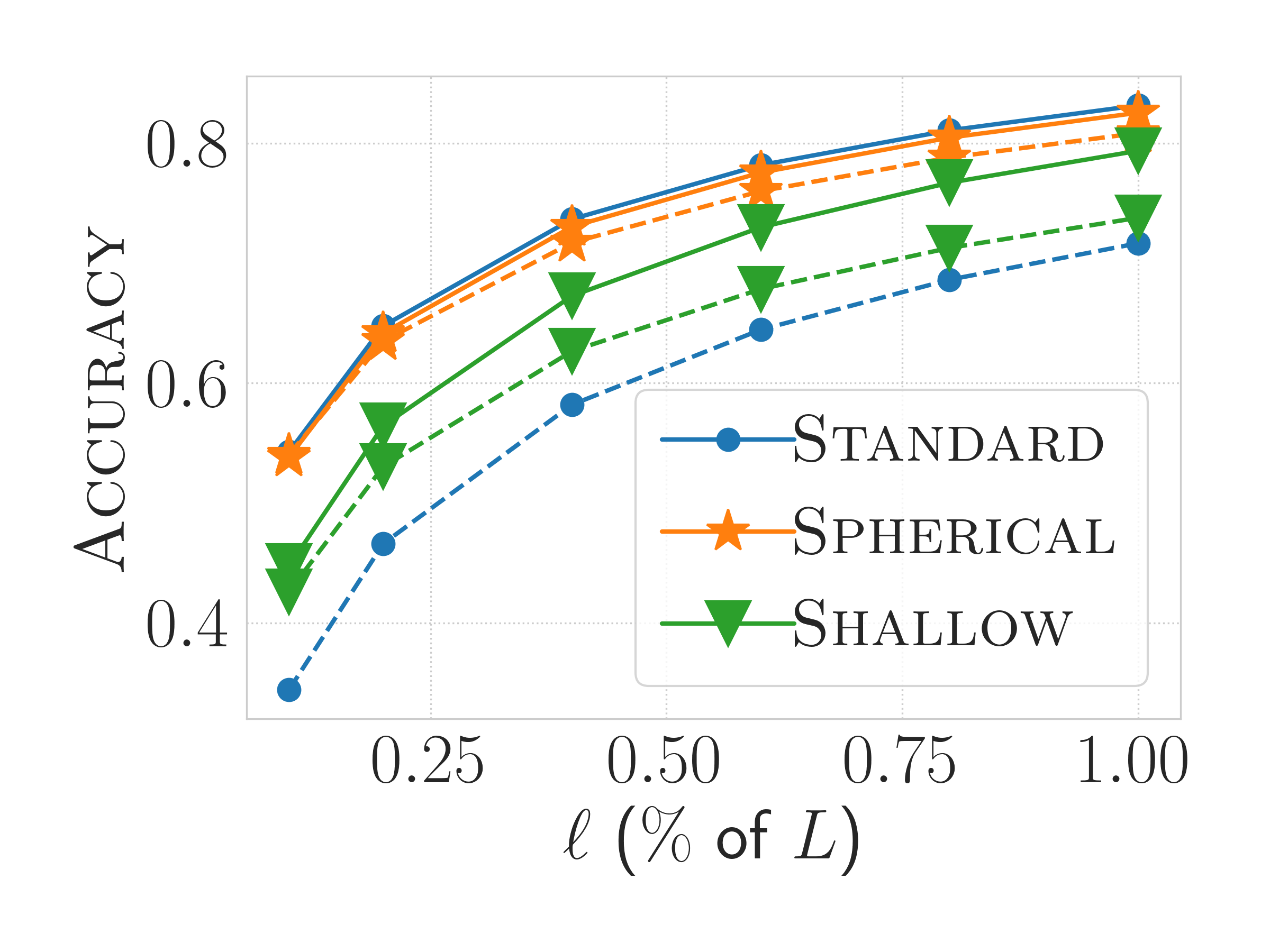}
        \caption{\footnotesize \textsc{MS MARCO}}
        \label{fig-a:top-100-msmarco}
    \end{subfigure}
    \hfill
    \begin{subfigure}[b]{0.325\textwidth}
        \centering
        \includegraphics[width=\textwidth]{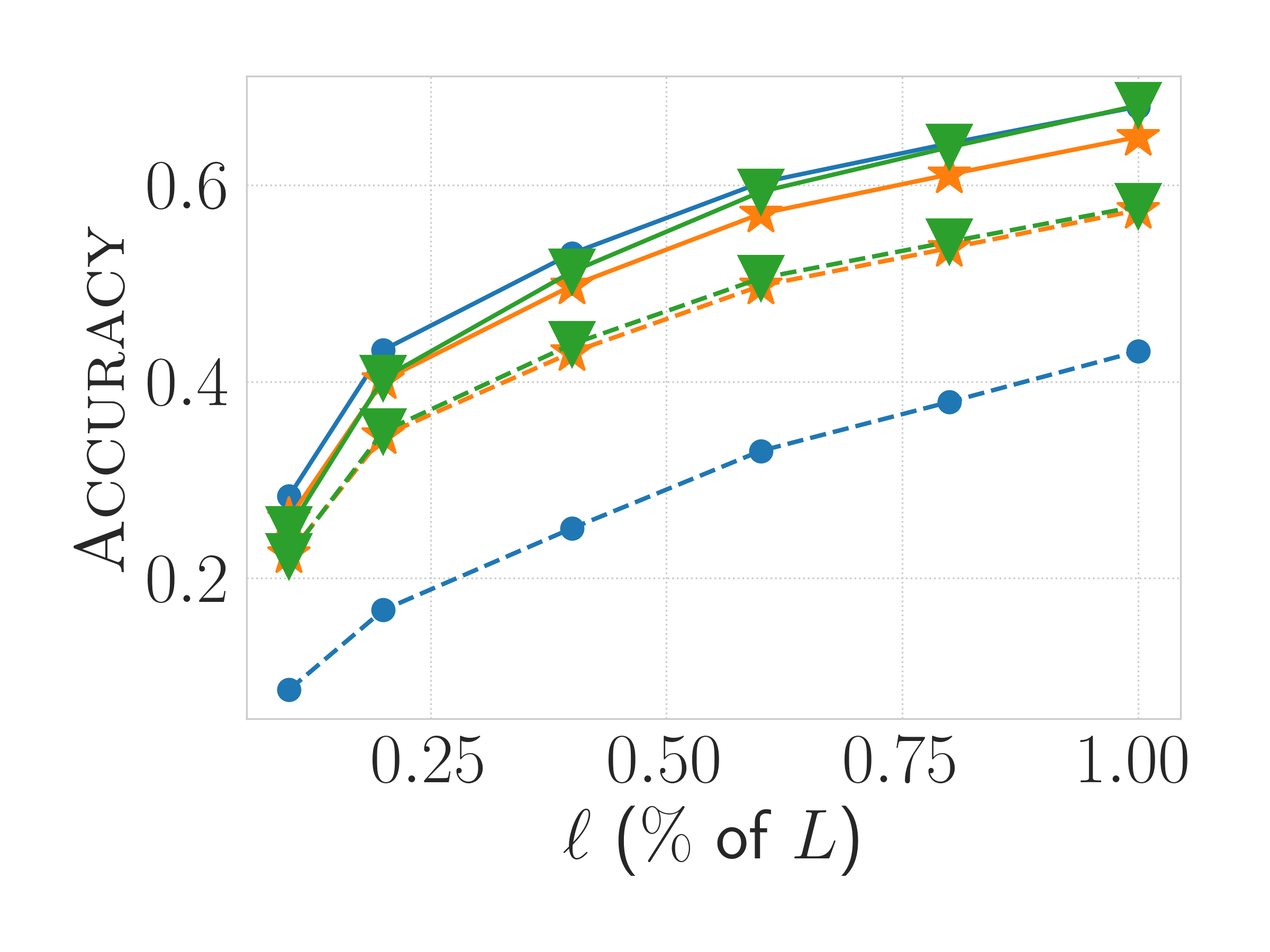}
        \caption{\footnotesize \textsc{HotpotQA}}
        \label{fig-b:top-100-hotpotqa}
    \end{subfigure}
    \hfill
    \begin{subfigure}[b]{0.325\textwidth}
        \centering
        \includegraphics[width=\textwidth]{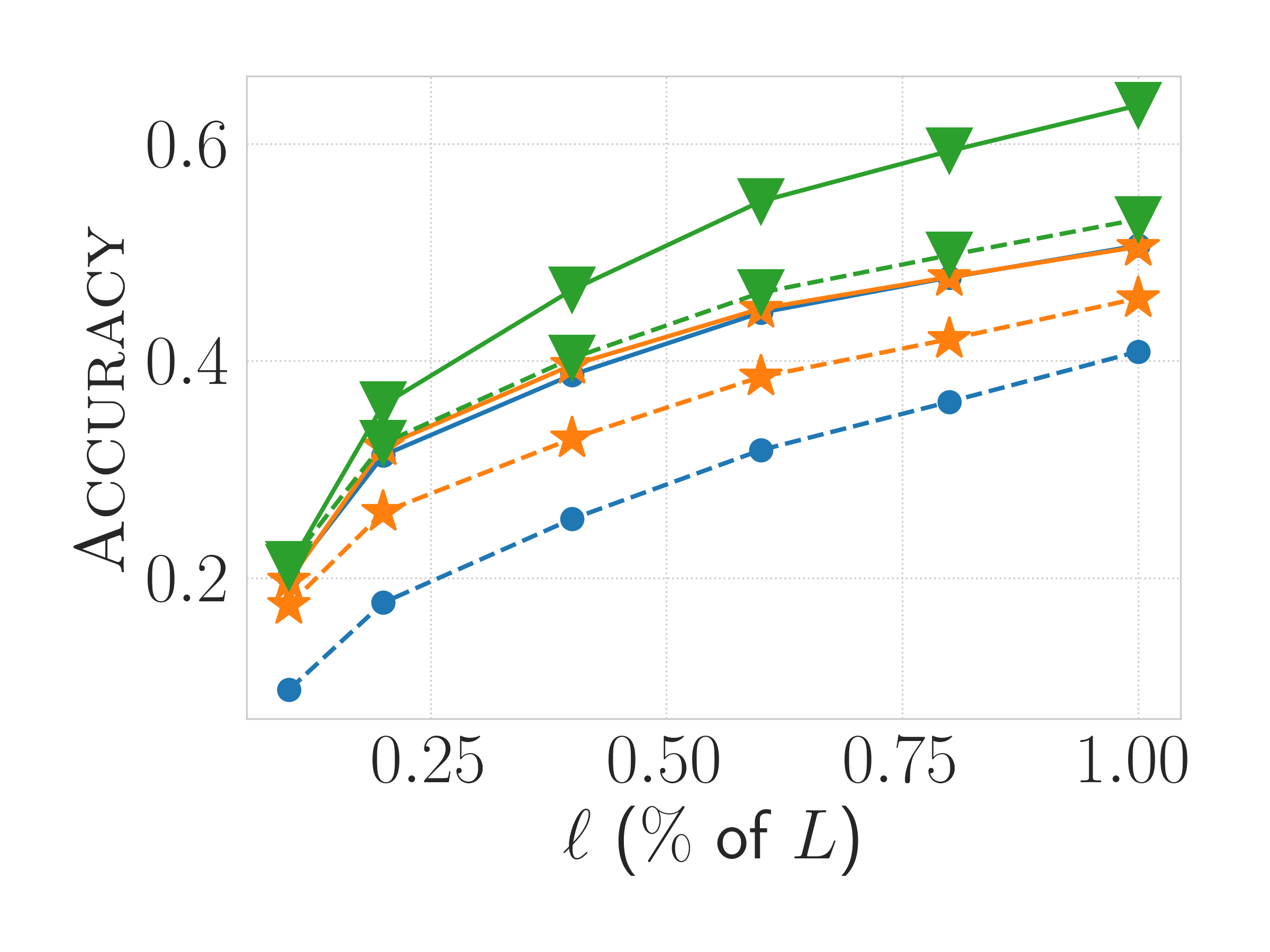}
        \caption{\footnotesize \textsc{FEVER}}
        \label{fig-c:top-100-fever}
    \end{subfigure}
    \caption{\footnotesize
    Top-$100$ accuracy of Learnt and 
    Baseline methods on \textsc{MS MARCO}, 
    \textsc{HotpotQA}, and \textsc{FEVER} 
    datasets using the \texttt{all-MiniLM-L6-v2} 
    embedding model. The $x$-axis 
    represents $\ell$, expressed as a 
    percentage of $L$,  while the 
    $y$-axis represents the accuracy.
    Solid lines indicate the Learnt 
    method, and dashed lines indicate
    the Baseline method.
    }
    \label{fig:top-100-results}
\end{figure}

Until now, our analysis has focused on 
top-$1$ accuracy. We now turn our attention 
to top-$k$ accuracy, with $k > 1$.
Figures~\ref{fig:top-10-results} and 
\ref{fig:top-100-results} depict the 
obtained results for top-$10$ and 
top-$100$ accuracy, respectively, 
as a function of $\ell$.

The findings discussed in the previous 
subsection also hold for top-$k$ 
accuracy with $k > 1$: the Learnt method
demonstrates superior performance 
compared to the Baseline, especially for 
smaller values of $\ell$; and, 
the accuracy gap between the two methods 
is most pronounced for Standard, followed 
by Shallow, and finally, Spherical KMeans.

These results are particularly 
interesting given that our proposed 
method is specifically designed to 
optimize top-$1$ accuracy, where the goal
is to identify the cluster associated with
the top-$1$ document for a given query. 
Despite this, as is evident from 
Figures~\ref{fig:top-10-results} and 
\ref{fig:top-100-results}, our model demonstrates 
remarkable performance for top-$k$
accuracy as well.

These results, obtained for top-$k$ 
accuracy, are highly promising and 
point to a potential avenue for 
generalizing the loss function to 
the top-$k$ scenario, as detailed in 
Section 5.2.1, with the potential for 
further improved accuracy.
A more in-depth exploration of this 
generalization is planned for future 
work.

\section{Further Analysis}

Our objective was to learn a linear 
routing function, $\acute{\tau}$, to 
obtain representative points that best
capture the content of the clusters.
This allowed us to ensure that when a 
new query $q$, arrived, the inner product 
between the query and the learnt 
representative points would more accurately 
reflect the similarity between $q$ and 
the clusters. This, in turn, led to 
improved accuracy in retrieving the top-$1$ 
document or, more generally, the top-$k$
documents.

Let us now consider a scenario where our 
goal is not to learn new representatives 
but to focus solely on learning a routing 
function that, given a query, returns a 
score for each cluster corresponding to 
the likelihood of finding the top-$1$  
document within that cluster.
$\acute{\tau}$ emerges as a linear 
solution to this problem. However, what 
happens if we relax the assumption of 
linearity? By studying a nonlinear routing
function, $\tilde{\tau}$, we will explore 
the impact of nonlinearity on solving 
this problem.

\subsection{Nonlinearity}

\begin{figure}
    \centering
    \hspace{2cm}
    \begin{subfigure}[b]{0.325\textwidth}
        \includegraphics[width=\textwidth]{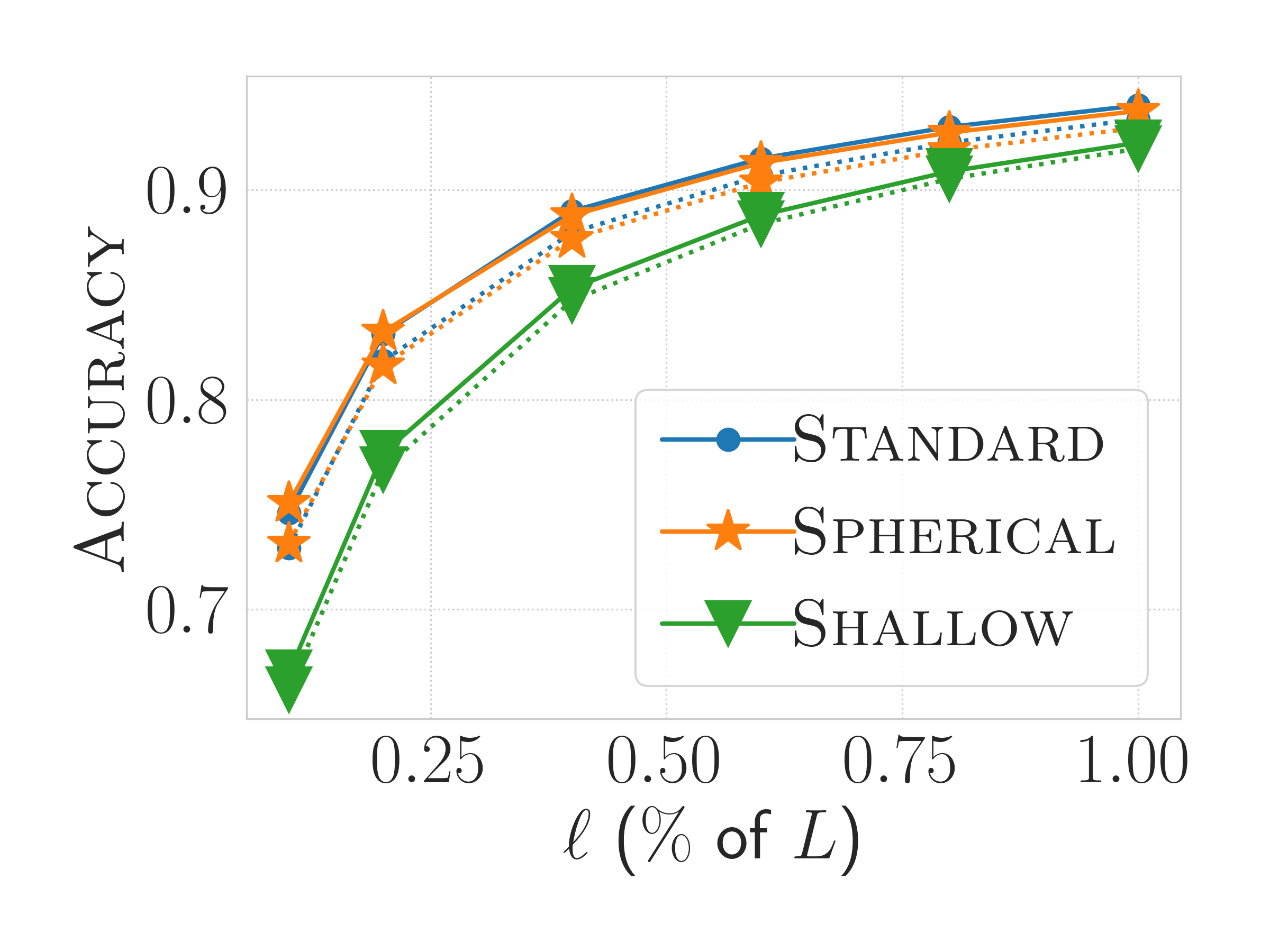}
        \caption{\footnotesize $\acute{\tau}$ vs. $\tilde{\tau}_{512}$}
        \label{fig-a:analysis-top-1-512}
    \end{subfigure}
    \hfill
    \begin{subfigure}[b]{0.325\textwidth}
        \includegraphics[width=\textwidth]{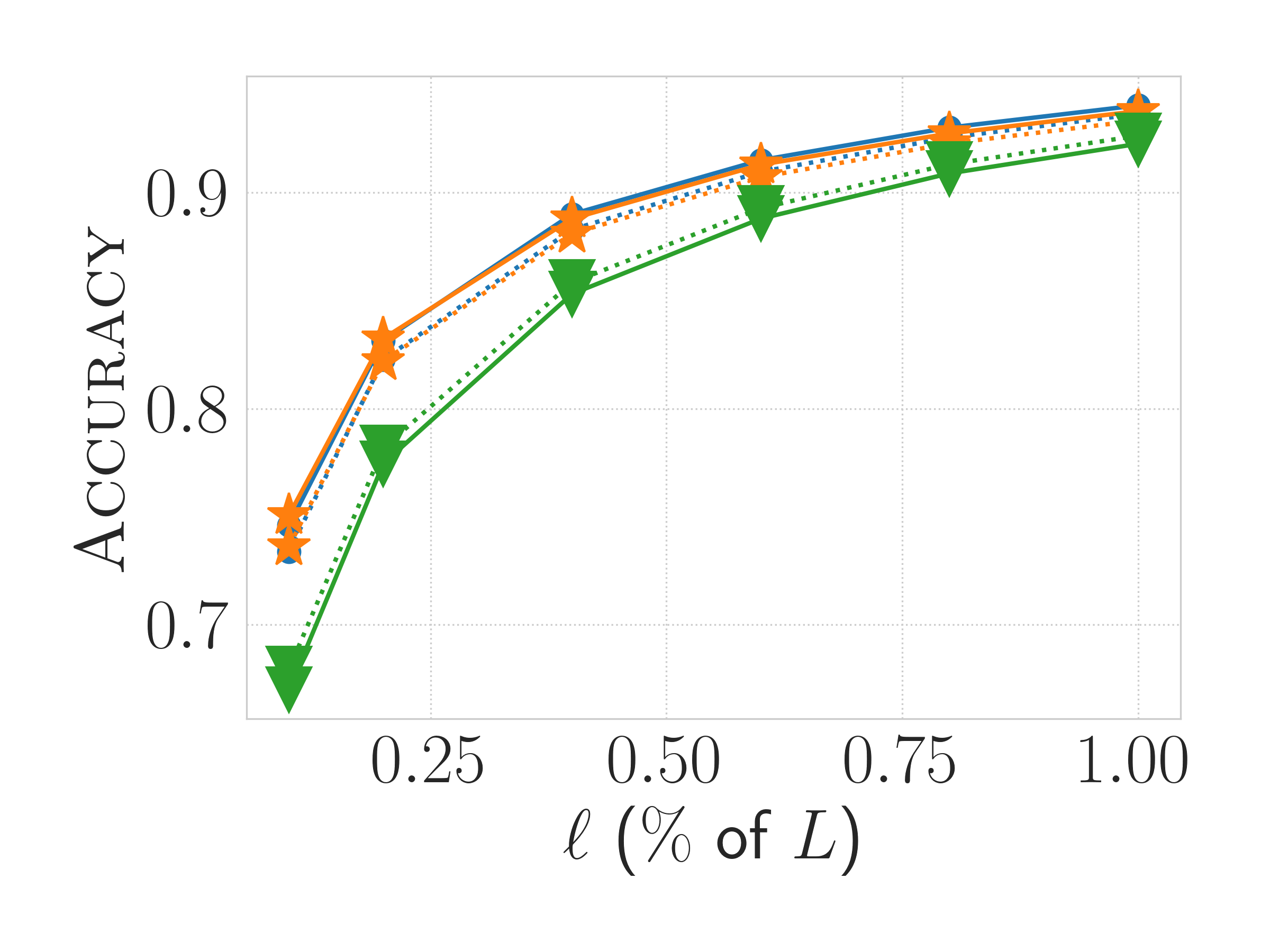}
        \caption{\footnotesize $\acute{\tau}$ vs. $\tilde{\tau}_{1024}$}
        \label{fig-b:analysis-top-1-1024}
    \end{subfigure}
    \hspace{2cm}

    \hspace{2cm}
    \begin{subfigure}[b]{0.325\textwidth}
        \includegraphics[width=\textwidth]{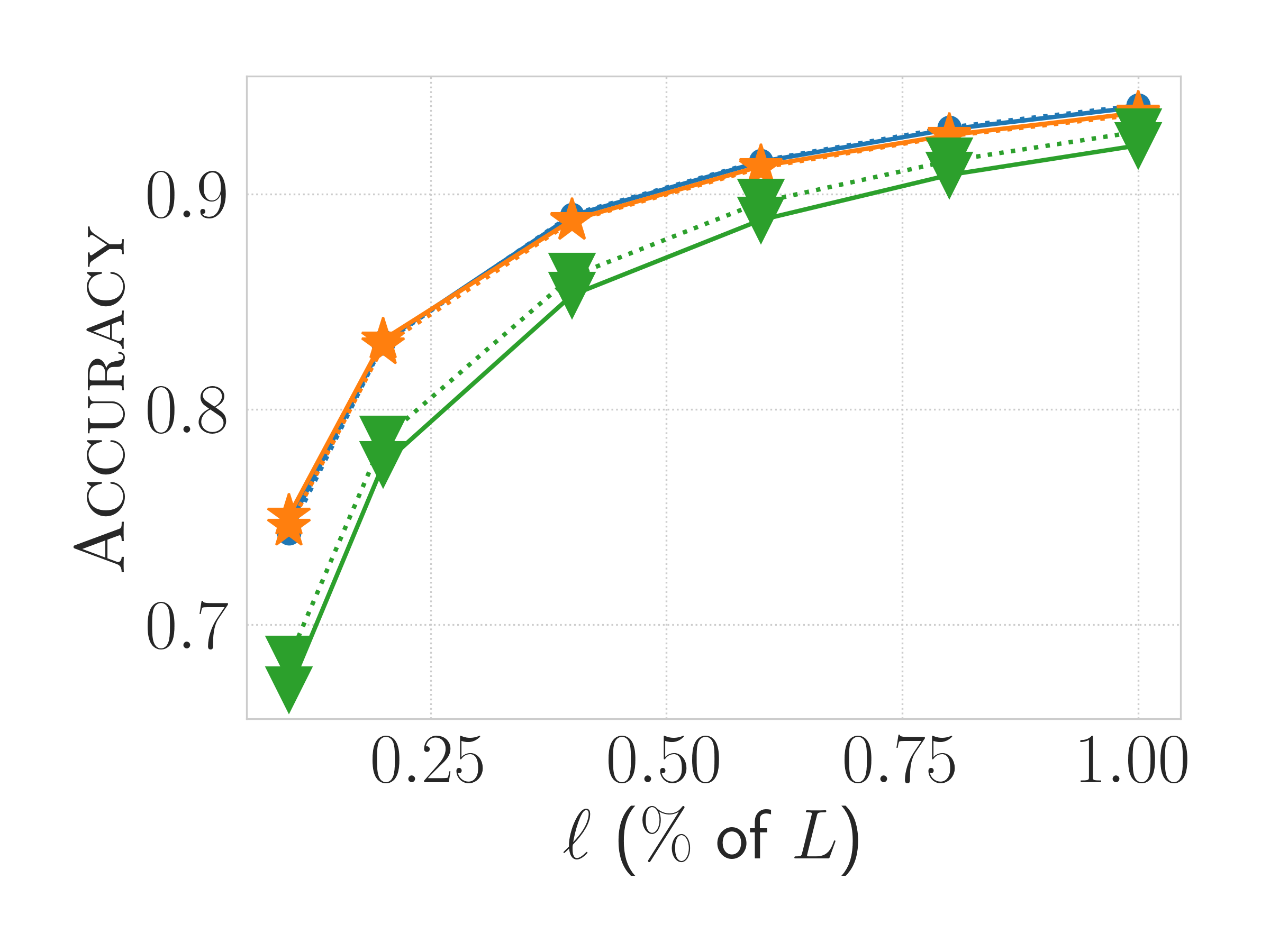}
        \caption{\footnotesize $\acute{\tau}$ vs. $\tilde{\tau}_{2048}$}
        \label{fig-c:analysis-top-1-2048}
    \end{subfigure}
    \hfill
    \begin{subfigure}[b]{0.325\textwidth}
        \includegraphics[width=\textwidth]{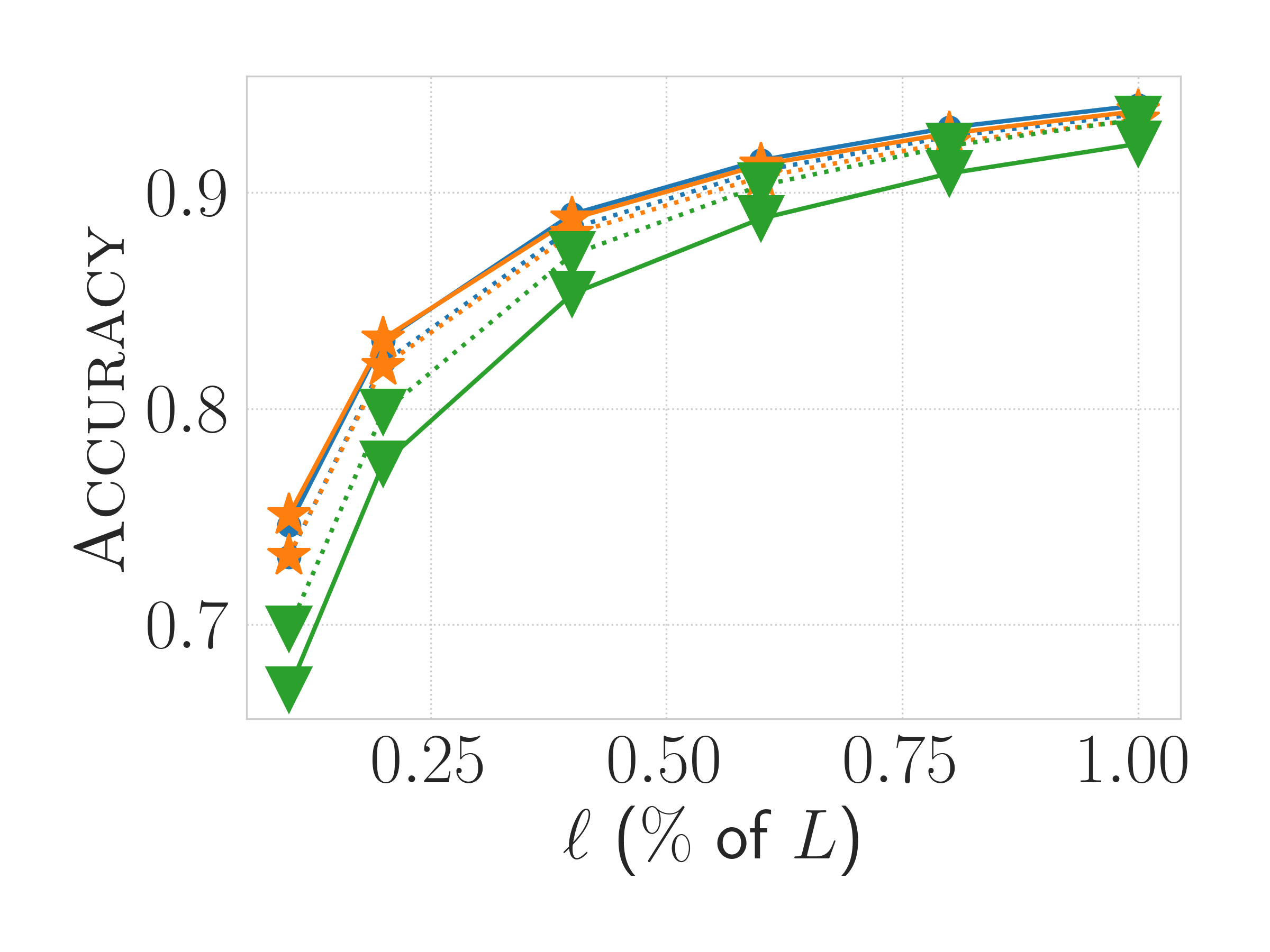}
        \caption{\footnotesize $\acute{\tau}$ vs. $\tilde{\tau}_{4096}$}
        \label{fig-d:analysis-top-1-4096}
    \end{subfigure}
    \hspace{2cm}
    \caption{\footnotesize
    Top-1 accuracy of the learnt linear 
    routing function, $\acute{\tau}$, 
    and the learnt nonlinear routing 
    function, $\tilde{\tau}$, on the 
    \textsc{MS MARCO} dataset using the 
    \texttt{all-MiniLM-L6-v2} embedding 
    model. The $x$-axis represents $\ell$ 
    as a percentage of $L$, while the 
    $y$-axis represents the accuracy. 
    Solid lines indicate $\acute{\tau}$, 
    and dotted lines indicate 
    $\tilde{\tau}$ variants with 
    different hidden layer sizes:
    $512$ (a), $1024$ (b), $2048$ (c),
    and $4096$ (d).
    }
    \label{fig:analysis-top-1-nonlinear}
\end{figure}

We explore and analyze the use of 
nonlinearity to solve the routing 
problem by defining a function, 
$\tilde{\tau}$, and comparing it 
to the linear routing function, 
$\acute{\tau}$.

$\tilde{\tau}$ is defined as a neural 
network with a structure similar to 
$\acute{\tau}$, but with some 
modifications: it includes a hidden layer,
employs the ReLU activation 
function~\cite{Agarap_2019_ReLU_paper}
in this layer, and incorporates biases 
in its neurons.
For our analyses, we specifically consider 
four variants of $\tilde{\tau}$: one with 
$512$ units in the hidden layer 
($\tilde{\tau}_{512}$), one with $1024$ units 
($\tilde{\tau}_{1024}$), one with $2048$ units 
($\tilde{\tau}_{2048}$), and one with 
$4096$ units ($\tilde{\tau}_{4096}$).

\begin{figure}
    \centering
    \hspace{2cm}
    \begin{subfigure}[b]{0.325\textwidth}
        \includegraphics[width=\textwidth]{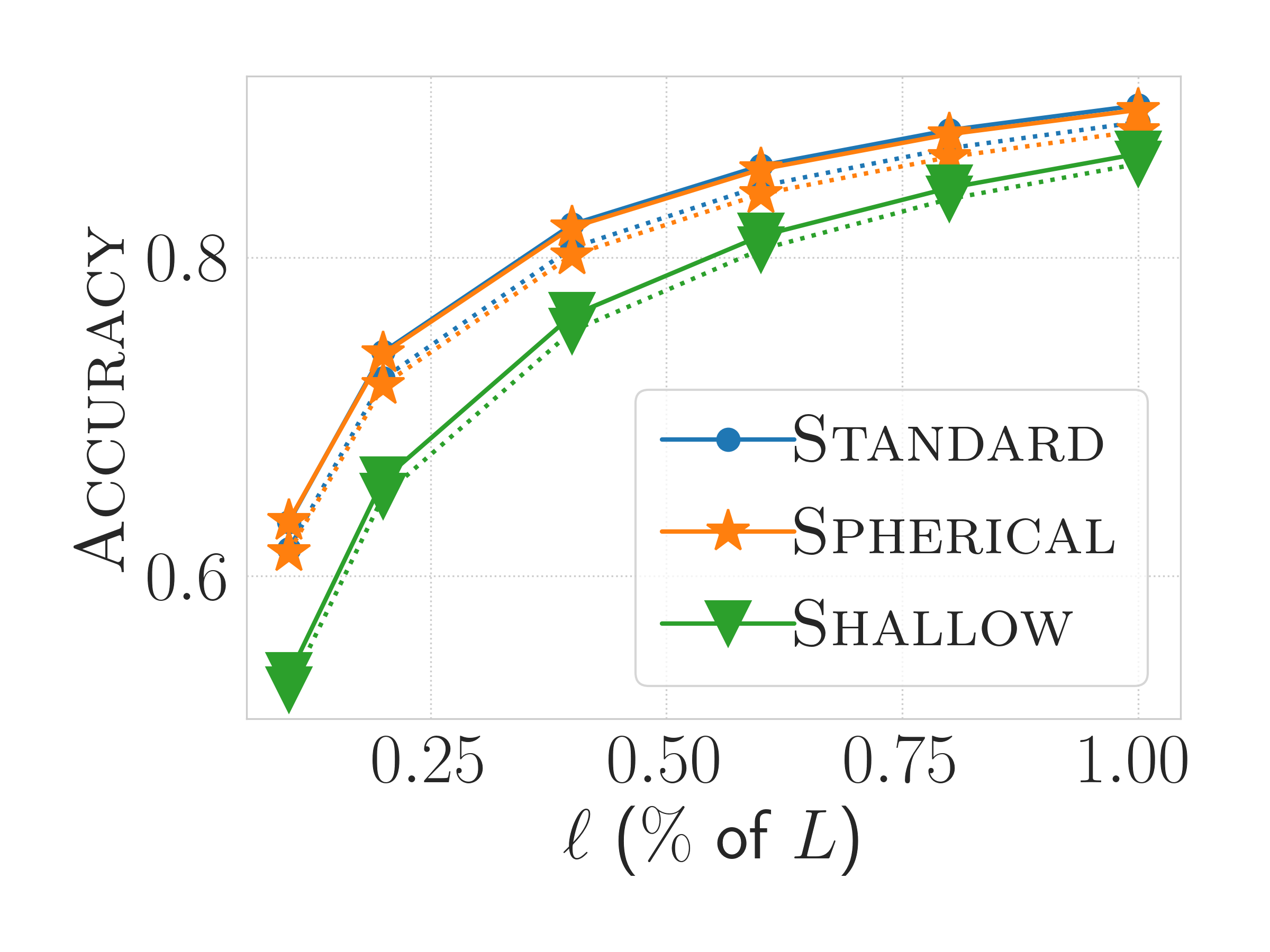}
        \caption{\footnotesize $\acute{\tau}$ vs. $\tilde{\tau}_{512}$}
        \label{fig-a:analysis-top-10-512}
    \end{subfigure}
    \hfill
    \begin{subfigure}[b]{0.325\textwidth}
        \includegraphics[width=\textwidth]{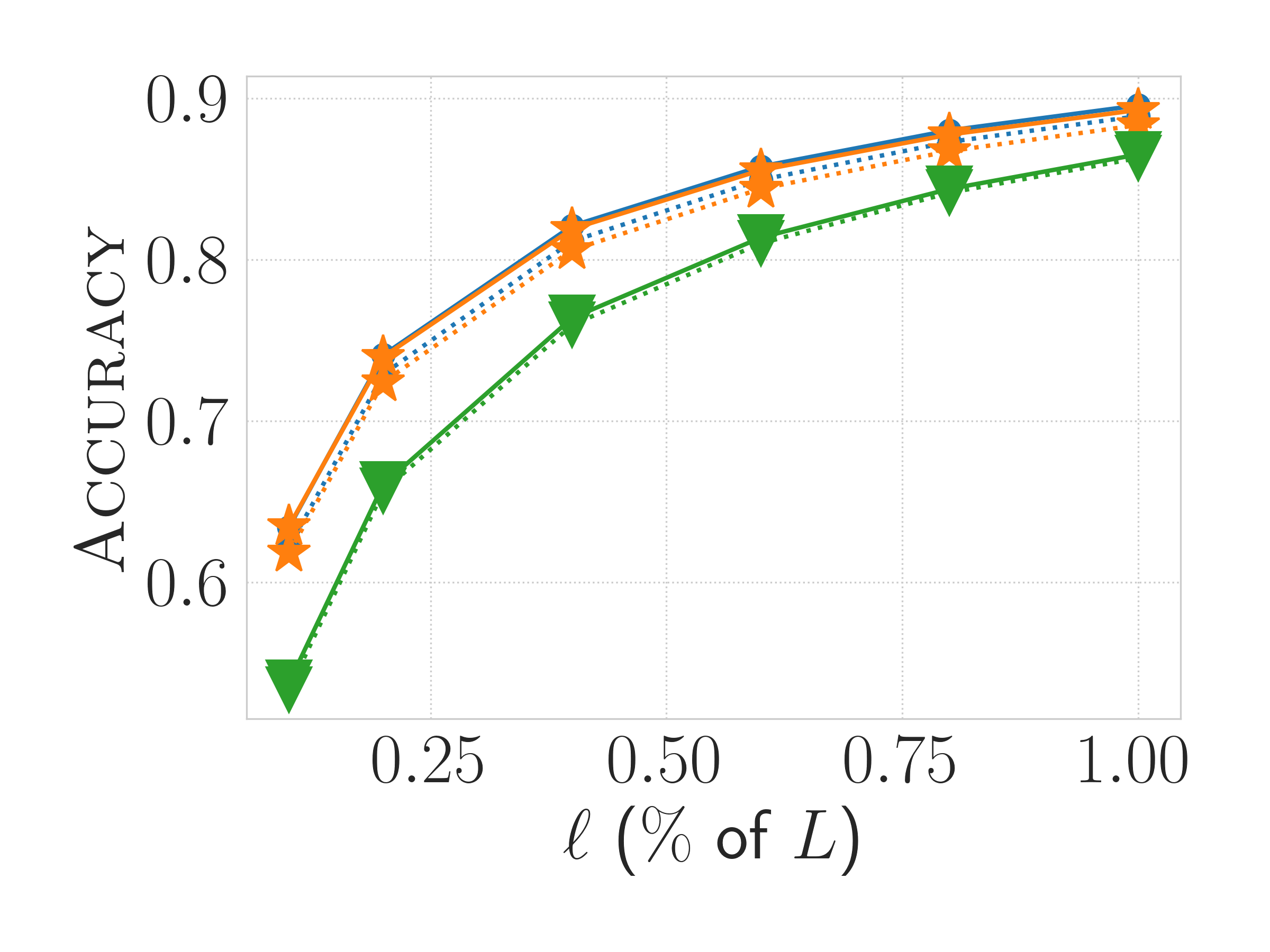}
        \caption{\footnotesize $\acute{\tau}$ vs. $\tilde{\tau}_{1024}$}
        \label{fig-b:analysis-top-10-1024}
    \end{subfigure}
    \hspace{2cm}

    \hspace{2cm}
    \begin{subfigure}[b]{0.325\textwidth}
        \includegraphics[width=\textwidth]{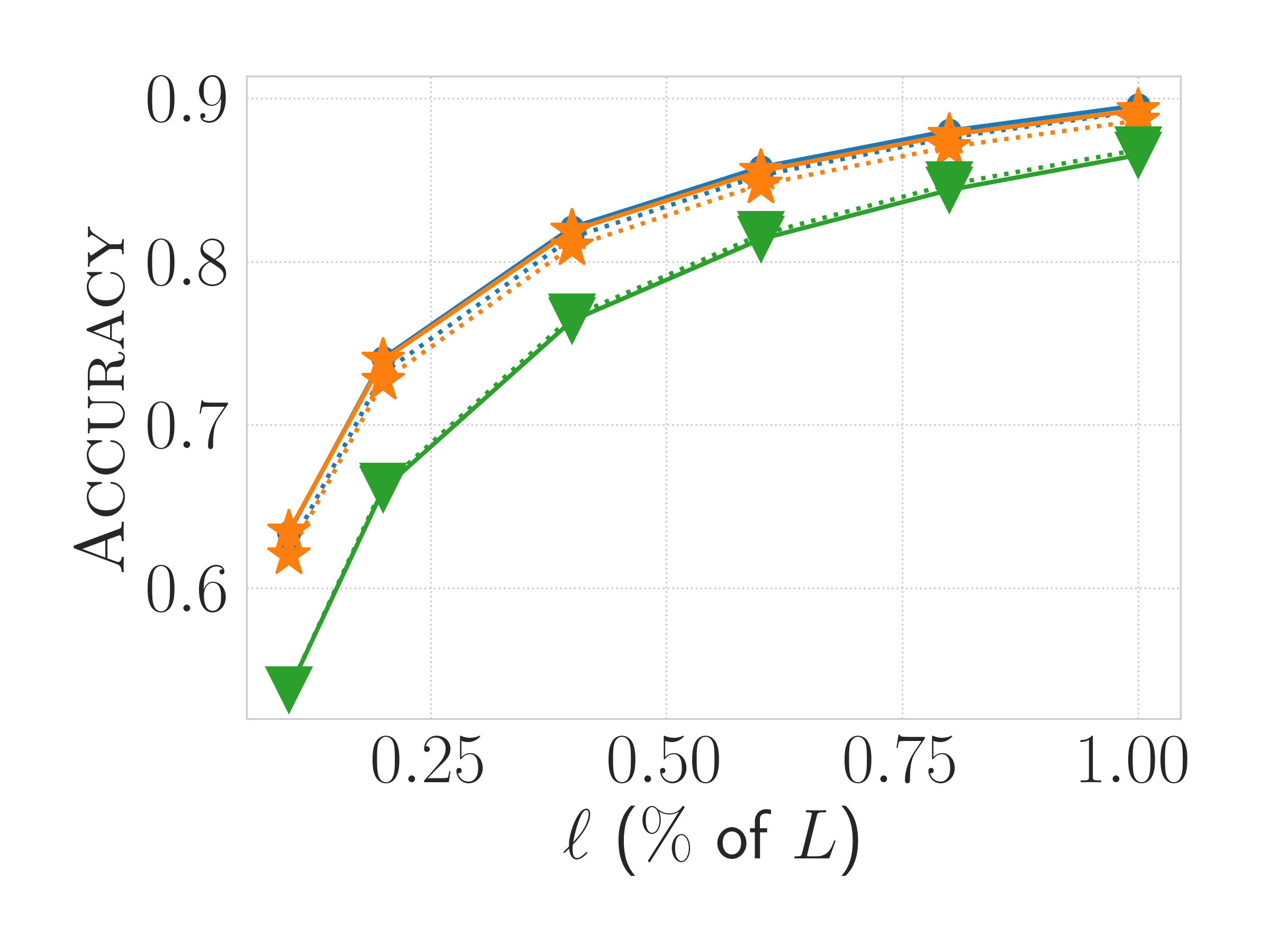}
        \caption{\footnotesize $\acute{\tau}$ vs. $\tilde{\tau}_{2048}$}
        \label{fig-c:analysis-top-10-2048}
    \end{subfigure}
    \hfill
    \begin{subfigure}[b]{0.325\textwidth}
        \includegraphics[width=\textwidth]{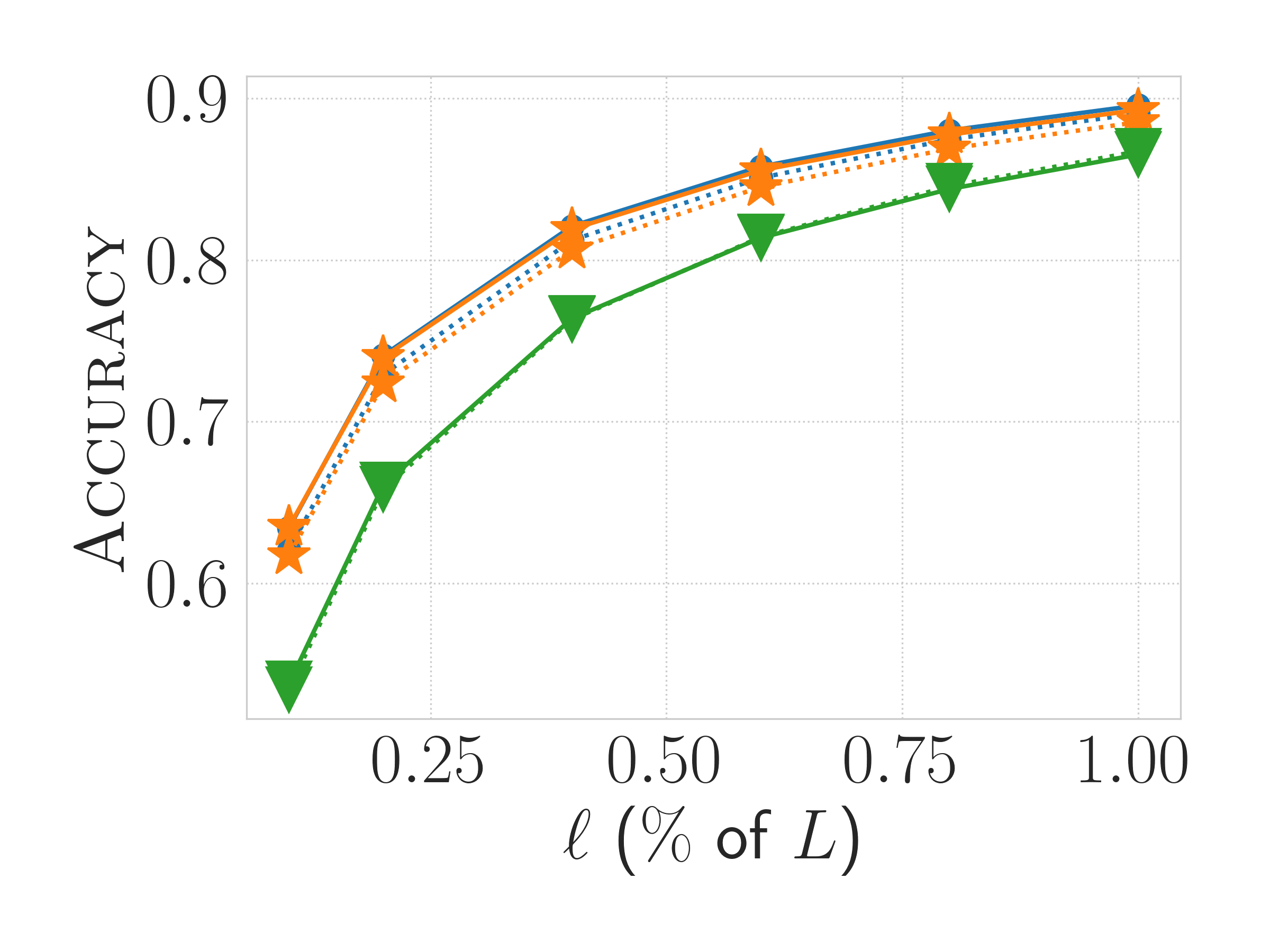}
        \caption{\footnotesize $\acute{\tau}$ vs. $\tilde{\tau}_{4096}$}
        \label{fig-d:analysis-top-10-4096}
    \end{subfigure}
    \hspace{2cm}
    \caption{\footnotesize
    Top-10 accuracy of the learnt linear 
    routing function, $\acute{\tau}$, 
    and the learnt nonlinear routing 
    function, $\tilde{\tau}$, on the 
    \textsc{MS MARCO} dataset using the 
    \texttt{all-MiniLM-L6-v2} embedding 
    model. The $x$-axis represents $\ell$ 
    as a percentage of $L$, while the 
    $y$-axis represents the accuracy. 
    Solid lines indicate $\acute{\tau}$, 
    and dotted lines indicate 
    $\tilde{\tau}$ variants with 
    different hidden layer sizes:
    $512$ (a), $1024$ (b), $2048$ (c),
    and $4096$ (d).
    }
    \label{fig:analysis-top-10-nonlinear}
\end{figure}

Figure~\ref{fig:analysis-top-1-nonlinear}
illustrates the top-$1$ accuracy for 
$\acute{\tau}$ and $\tilde{\tau}$ as a 
function of $\ell$, where $\ell$ varies
between $0.1\% \times L$ and $1\% \times L$,
on \textsc{MS MARCO} dataset using 
\texttt{all-MiniLM-L6-v2} as embedding model.

Figure~\ref{fig:analysis-top-1-nonlinear} 
shows that, for the nonlinear routing 
function, considering both Standard and 
Spherical KMeans clustering algorithms, 
the best results are obtained with the 
$\tilde{\tau}_{2048}$ variant.

Focusing now on the comparison between 
$\tilde{\tau}_{2048}$ and $\acute{\tau}$ 
in terms of Standard and Spherical KMeans,
it can be observed that despite the 
nonlinearity of $\tilde{\tau}_{2048}$, 
the two routing functions exhibit nearly
identical accuracy results.
Specifically, for lower values of $\ell$, 
$\acute{\tau}$ outperforms $\tilde{\tau}_{2048}$ 
by almost $1\%$.

For the Shallow KMeans clustering algorithm, 
however, the best performing variant of 
$\tilde{\tau}$ is $\tilde{\tau}_{4096}$.
$\tilde{\tau}_{4096}$ achieves higher accuracy 
values compared to $\acute{\tau}$, 
with improvements of approximately $4.2\%$ for
$\ell = L / 1000$ and $1.2\%$ for 
$\ell = L / 100$.

Analysis of the results indicates that 
nonlinearity does not provide significant 
improvements in terms of top-$1$ accuracy.
Consequently, the learnt linear routing 
function, $\acute{\tau}$, emerges as the 
optimal choice over the nonlinear 
functions examined. This conclusion is 
supported by the higher accuracy values 
achieved by $\acute{\tau}$ in Spherical  
and Standard KMeans clustering algorithms, 
as well as by the fact that the additional 
complexity associated with $\tilde{\tau}_{4096}$ 
in Shallow KMeans does not justify the 
relatively small improvement in accuracy.
Considering only the number of parameters, 
$\tilde{\tau}_{4096}$ has $13{,}867{,}960$ 
parameters, significantly more than the 
$1{,}155{,}000$ parameters of $\acute{\tau}$.

Similar trends can be observed for the 
top-$10$ accuracy, as depicted in 
Figure~\ref{fig:analysis-top-10-nonlinear}.
The only exception pertains to Shallow 
KMeans: the best results are achieved with 
the $\tilde{\tau}_{2048}$ variant rather 
than $\tilde{\tau}_{4096}$; moreover, the 
accuracy difference between $\tilde{\tau}_{2048}$ 
and $\acute{\tau}$ is narrower compared to 
the previous case. This further supports 
the argument in favor of $\acute{\tau}$ 
over $\tilde{\tau}$.

In conclusion, in light of the findings 
presented in this section, it can be 
concluded that the learnt linear routing 
function, $\acute{\tau}$, is the most 
suitable choice compared to the nonlinear 
functions examined. $\acute{\tau}$ offers 
the best balance of efficiency and 
effectiveness that is essential in the 
field of ANN search.
\chapter{Conclusions and Future Work}

In this work, we have presented a new 
state-of-the-art for clustering-based 
ANN search. Our proposed approach 
involves learning a linear routing function 
via LTR to generate learnt representative 
points. These points provide a more 
accurate and semantically rich 
representation of each cluster,
significantly improving ANN search 
accuracy.

To facilitate a comprehensive 
understanding of our method, we first have
provided a detailed overview of 
clustering-based ANN search, starting from
the fundamental concepts such as vector 
search, ANN search, and clustering.
Subsequently, we have introduced LTR, 
focusing on the essential ingredients 
for learning a ranking function.

Having established a solid foundation, 
encompassing all the necessary elements 
for a thorough understanding of our method,
we then have proceed to present our approach 
in detail, first providing an intuitive 
explanation and then a formal definition.

Our proposed method is grounded in an 
intuitive observation: The routing function, 
which determines the most promising clusters 
for a given query, solves a ranking problem.
In other words, the routing function ranks 
clusters based on their relevance to the 
query, specifically by ranking them 
according to their probability of containing 
the query's nearest neighbor.

This insight has led us to an innovative 
solution: Learning the routing function
via LTR. In particular, we learn a simple 
linear function where the parameters of the 
function represent the new representative 
points for each cluster, more informative 
and discriminative than the standard 
representative points.
It is noteworthy how straightforward 
it has been to prepare the necessary 
ingredients for learning the routing 
function, from the dataset to the loss
function.

Through extensive experiments conducted  
on diverse datasets, embedding models, 
and clustering algorithms, we have 
empirically demonstrated that our 
proposed method consistently outperforms 
the baseline, significantly improving the 
accuracy of MIPS-based clustering-based 
ANN search.
The learnt representative points, compared 
to standard representative points, enable 
the retrieval of top-$\ell$ partitions 
containing the top-$1$ or top-$k$ documents 
for a given query with significantly 
higher accuracy.

Furthermore, we have conducted additional 
investigations into learning the routing 
function by relaxing the linearity 
assumption and learning a nonlinear routing 
function, with the sole purpose of returning
the top-$\ell$ partitions for a given query.
Our findings have revealed that the learnt 
linear routing function emerged as the optimal 
choice, offering the best balance between 
efficiency and effectiveness compared to the 
learnt nonlinear routing functions analyzed.

In summary, the proposed method emerges as a 
conceptually simple and the state-of-the-art 
approach for clustering-based ANN search.
Moreover, by developing this method, we have 
demonstrated the potential of integrating two 
important fields of IR, LTR and ANN.
Additionally, we have showcased the potential 
of learning representative points for groups 
of elements, an idea that can be extended to 
various domains.

Equally exciting is the fact that this work 
has also raised numerous questions for future 
research. 
These include generalizing the learning 
approach for the routing function to the 
top-$k$ problem, with $k > 1$, as discussed in 
Section 5.2.1; extending the method to other 
distance functions such as Euclidean distance, 
cosine distance, or Manhattan distance; 
investigating the influence of vector 
dimensionality of queries and documents on the 
accuracy of our method; and finally, exploring 
the new area of query-aware clustering for ANN 
search, where the document space is clustered 
using the query dataset.

\addcontentsline{toc}{chapter}{Bibliography}
\bibliographystyle{plain}
{\footnotesize
 \bibliography{biblio}}

\end{document}